\documentclass[%
 amsmath,amssymb,
 aps,
]{revtex4-2}

\usepackage{graphicx}% Include figure files
\usepackage{dcolumn}% Align table columns on decimal point
\usepackage{bm}
\usepackage{subcaption}
\usepackage{xcolor}
\usepackage[normalem]{ulem}
\newcommand{\stkout}[1]{\ifmmode\text{\sout{\ensuremath{#1}}}\else\sout{#1}\fi}
\usepackage{cancel}
\DeclareMathOperator{\sgn}{sgn}

\newcommand{\e}{\epsilon}

\newcommand{\om}{\omega}

\newcommand{\pa}{\partial}

\newcommand{\al}{\alpha}
\newcommand{\gam}{\gamma}

\newcommand{\da}{\dagger}
\newcommand{\nab}{ {\bf \nabla} }
\newcommand{\bu}{ {\bf u} }
\newcommand{\bU}{ {\bf U} }
\newcommand{\hbu}{ \hat{\bf u} }

\newcommand{\ba}{ {\bf a} }
\newcommand{\bb}{ {\bf b} }
\newcommand{\bx}{ {\bf x} }
\newcommand{\hxi}{ \hat{\xi} }
\newcommand{\bff}{ {\bf f} }

\newcommand{\bq}{ {\bf q} }
\newcommand{\di}{\mathrm{d}}
\newcommand{\Ren}{\text{Re}}
\newcommand{\Renc}{\text{Re}_{\text{c}}}
\newcommand{\Her}{\text{H}}
\newcommand{\tc}{\text{c}}
\newcommand{\tq}{\text{q}}
\newcommand{\ys}{y_{\text{s}}}
\newcommand{\ti}{\text{i}}

\newcommand{\buR}{{\bf u}_{2}^{A^2}}
\newcommand{\buV}{{\bf u}_{2}^{\Delta}}
\newcommand{\pR}{p_{2}^{A^2}}
\newcommand{\pV}{p_{2}^{\Delta}}

\newcommand{\PS}{\bar{P}}
\newcommand{\PSw}{\bar{P}^{\text{w}}}

\usepackage{mathtools}

\newcommand\Fo[1]{\mathcal{F} \left [ #1 \right]}
\newcommand\Foi[1]{\mathcal{F}^{-1} \left [ #1 \right]}
\newcommand\Fos[2]{\mathcal{F}_{ #1 }\left [ #2 \right]}

\newcommand\ssp[2]{\left \langle #1 \middle | #2 \right \rangle}

\newcommand\mat[2]{\begin{bmatrix} #1 \\ #2 \end{bmatrix}}

\newcommand \ea[1]{\langle #1 \rangle }

\begin{document}

\preprint{APS/123-QED}

\title{Noise-induced transitions past the onset of a steady symmetry-breaking bifurcation: the case of the sudden expansion}% Force line breaks with \\

\author{Yves-Marie Ducimeti\`{e}re}%
\email{yves-marie.ducimetiere@epfl.ch}
\author{Edouard Boujo}
\author{Fran\c cois Gallaire}
\affiliation{%
 Laboratory of Fluid Mechanics and Instabilities, \'Ecole Polytechnique F\'ed\'erale de Lausanne, Lausanne CH-1015, Switzerland
}%

\date{\today}% It is always \today, today,

\begin{abstract}
We consider fluid flows, governed by the Navier-Stokes equations, subject to a steady symmetry-breaking bifurcation and forced by a weak noise acting on a slow time scale. By generalizing the multiple-scale weakly nonlinear expansion technique employed in the literature for the response of the Duffing oscillator, we rigorously derive a stochastically forced Stuart-Landau equation for the dominant symmetry-breaking mode. The probability density function of the solution, and of the escape time from one attractor to the other, are then determined by solving the associated Fokker-Planck equation. The validity of this reduced order model is tested on the flow past a sudden expansion, for a given Reynolds number and different noise amplitudes. At a very low numerical cost, the statistics obtained from the amplitude equation accurately reproduce those of long-time direct numerical simulations.
\end{abstract}

\maketitle

%\tableofcontents
%%%%%%%%%%%%%%%%%%%%%%%%%%%%%%%%%%%%%%%%%%
%%%%%%%%%%%%%%%%%%%%%%%%%%%%%%%%%%%%%%%%%%
%%%%%%%%%%%%%%%%%%%%%%%%%%%%%%%%%%%%%%%%%%
%%%%%%%%%%%%%%%%%%%%%%%%%%%%%%%%%%%%%%%%%%
\section{Introduction}
%%%%%%%%%%%%%%%%%%%%%%%%%%%%%%%%%%%%%%%%%%
%%%%%%%%%%%%%%%%%%%%%%%%%%%%%%%%%%%%%%%%%%
%%%%%%%%%%%%%%%%%%%%%%%%%%%%%%%%%%%%%%%%%%
%%%%%%%%%%%%%%%%%%%%%%%%%%%%%%%%%%%%%%%%%%

Fluid flows, governed by the Navier-Stokes equations, are known to exhibit an extremely rich phenomenology, including pattern formation, spatio-temporal chaos, turbulence, etc... Specifically, some fluid flows can reach an attracting coherent structure (e.g. fixed point) where they appear to be locally stationary for possibly extremely long times, but, from time to time, because of a rare fluctuation, exit the basin attraction of such coherent structure and transit towards another one. Although possibly restricted to some specific regimes in the parameter space, rare transitions are observed in completely different contexts. These include three- or two-dimensional experimental turbulent flows \cite{Sommeria86}, for instance, wakes behind an Ahmed body (a simplified car model) % for automobile bodies) 
\cite{Grandemange13} or an axisymmetric \cite{Brackston16,Callaham22} bluff body. 
Other examples are found in magneto-hydrodynamics experiments \cite{Berhanu07}, two-dimensional numerical turbulent flows \cite{Bouchet09, Bouchet19, Dallas20} and atmospheric flows \cite{Weeks97, Schmeits01}.

In the framework of equilibrium statistical mechanics, steady states of the system minimize a potential. Thenceforth, some laws describing the probability of rare transitions can be analytically derived. The Arrhenius law, in particular, considers a bi-stable overdamped system driven by a stochastic noise $\xi(t)$ at temperature $\Gamma_e$, $\mathrm{d}_t x = -\mathrm{d}_xV(x)  + \sqrt{2 k_B \Gamma_e}\xi(t)$, with $V$ a double-well potential of potential barrier $\Delta V$. The law predicts the expected time $\ea{\Delta T}$ between two transitions to scale like the exponential of minus the potential barrier separating two attractors, divided by the square of the fluctuations intensity, i.e. $\ea{\Delta T} \propto \exp(-\Delta V/( k_B \Gamma_e))$.  
 
However, fluid flows are mostly out-of-equilibrium systems, for energy is constantly injected (typically at boundaries) and dissipated by viscosity. In addition, Navier-Stokes equations have a continuously infinite number of degrees of freedom, which, after discretization, translates into thousands or millions of degrees of freedom, depending on the flow complexity. 
%Due to the out-of-equilibrium nature of most fluid flows, the route over the barrier from one basin to another is generally not the time reversal of the route from the latter to the former, implying that the space of reaction coordinates can't be reduced to one dimension. No potential then exist, and 
Therefore, characterizing rare transitions from one basin of attraction of the Navier-Stokes equations to another is a current scientific challenge.

There exist some specific numerical approaches in considering these transition events. All are motivated by the fact that a direct numerical simulation of the system is inappropriate in the statistical study of transition events, both because they occur over possibly extremely long time scales, and because the large number of degrees of freedom makes the numerical simulations of the Navier-Stokes equations unreasonably slow and energetically costly. Instead, the used numerical techniques incorporate some elements of the large deviation theory \cite{Freidlin98}, which considers non-equilibrium dynamical systems in the limit where they are forced by weak noise (consistent with the fact that transitions are rare). The general idea is to compute the most likely trajectory that links two given stable and distinct states. This specific trajectory is called an \textit{instanton}, and was shown in \cite{Freidlin98} to minimize an action in the path integral representation of the system. The probability of the transition along the instanton can also be computed and used to estimate the expected time between two transition events.

The instanton was numerically computed in different fluid flow modelizations, from two-dimensional geophysical turbulence \cite{Bouchet09, Bouchet11} to one-dimensional Burgers turbulence \cite{Grafke13}, to the transition between the plane Poiseuille flow and a traveling wave solution in a two-dimensional periodic domain \cite{Wan15}. More recently, \cite{Schorlepp22} considered the three-dimensional stochastically forced Navier–Stokes equations, and obtained the most likely configurations for extreme vorticity and strain events as the numerical solutions of the instanton problem. A more indirect manner to determine the instanton is by using the adaptive multilevel splitting algorithm \cite{Cerou07}, which is a rare event algorithm whose effect is to magnify the number of transition events. Thereby, very large statistics of transition paths can be produced, which are expected to concentrate around and reveal the instanton. Recently, this method was successfully employed in \cite{Bouchet19} for the turbulent flow obtained from a simplified model of Jupiter troposphere's dynamics. The work presented in \cite{Lestang20}, concerned about extreme mechanical forces exerted by a turbulent flow impinging on a bluff body, constitutes another example. It was also shown in a recent work \cite{Lecoanet18} that the computation of the instanton trajectory could be linked to a more basic nonlinear maximization problem of the flow kinetic energy \cite{Pringle10}.

To the knowledge of the authors, all the works mentioned so far relied on numerical tools. In the present paper, we shall employ a different strategy, expected to be valid at least in some simplified situations. In the spirit of \cite{Cates23}, who focused on the extension of the classical nucleation theory to an active phase-separating system, we first aim at analytically reducing the dimensionality of a flow forced by a weak and slowly varying stochastic noise. Then, analytical tools from statistical mechanics such as the Fokker-Planck equation will be deployed to compute the statistics of the reduced-order system, which is substantially easier to study and physically interpret than the original equation.

A general method for reducing the dimensionality of a nonlinear system was developed in dynamical system theory, and is valid asymptotically close to a bifurcation point in the parameter space, where an eigenmode of the linearized operator becomes unstable and grows exponentially. The idea is to derive an equation for the amplitude(s) of the bifurcating eigenmode(s). Such an equation is of minimal dimension (and nonlinear order) yet extracts the core of the nonlinear behavior of the original equation in the vicinity of a bifurcation point \cite{Guckenheimer83}. Its derivation relies on a clear multiple-scale asymptotic expansion procedure, and its first use in fluid mechanics dates back to \cite{Gor57} in the context of thermal convection. It was used in numerous studies since then and still is nowadays, for instance in \cite{Kerswell04, Sipp07, Shukla11, Ohm22, Buza22, Zampogna23} to cite only a few.

Of particular interest here, the analyses in \cite{Rusak99, Hawa01, Camarri19} derive a Stuart-Landau weakly nonlinear amplitude equation, for the steady symmetry-breaking eigenmode in the flow past a two-dimensional plane sudden expansion. Thereby, the two asymmetric attractors of the flow after the bifurcation could be approached as the equilibrium solutions of a single-degree-of-freedom equation, with good accuracy for Reynolds numbers asymptotically close to its critical value at the bifurcation. Note that a substantial body of work is devoted to the sudden expansion flow, due to its common appearance in the industrial or academic context. In addition to the ones already presented, could also be mentioned the studies in \cite{Lanzerstorfer12} and \cite{Debuysschere21}, concerned with the stability of the two-dimensional plane sudden expansion for different geometries, and inlet velocity profiles, respectively.

The construction of amplitude equations can be generalized to nonlinear dynamical systems subject to a stochastic forcing, as shown in \cite{Rajan88,Rong98} and \cite{Nayfeh90} for the Duffing and Duffing-Rayleigh oscillators, respectively. The inclusion of a noise term in an amplitude equation would have the effect of making its solution transit from one of its equilibrium solutions to another, for instance when the latter equation describes a supercritical or subcritical pitchfork or a subcritical Hopf bifurcation. The statistics of these transitions could be obtained with low computational efforts, yet in principle apply to the original equation, under the simplifying hypothesis made for the derivation of the amplitude equation.

However, if stochastic amplitude equations were found to be accurate models for some specific flows indeed \cite{Petrelis09, Brackston16, Callaham22}, the noise term systematically resulted from an \textit{ad-hoc} addition. In other words, the rigorous method deployed for the Duffing oscillators in \cite{Rajan88, Rong98} and \cite{Nayfeh90} to recover a noise term at the level of the amplitude equation directly from the stochastically forced original equation, does not seem to have been yet applied to the Navier-Stokes equations. Therefore, this will be the primary focus of the present paper.

For this purpose, we will consider a flow experiencing a supercritical pitchfork bifurcation, such that our method is a generalization of that outlined in \cite{Camarri19} and will result in a stochastically forced Stuart-Landau equation. Yet, the proposed method is expected to be adaptable to other fluid flows subject to multi-stability, thus to noise-induced transitions, closely after a bifurcation. Such situations include other flows experiencing a supercritical pitchfork bifurcation, such as the one in a pipe junction for some junction angles \cite{Chen17} or between a co-rotating disk pair for some gap ratio \cite{Randriamampianina01}. Supercritical pitchfork bifurcations also occur in laminar (or turbulent) three-dimensional wakes of rectangular prisms for some aspect ratios \cite{Zampogna23}, in the granular plane Couette flow for some parameters \cite{Shukla11}, in active suspensions of elongated swimming particles (immotile shakers or motile pullers/pushers) for some swimming speed \cite{Ohm22}, in the two-dimensional flow past an inverted flag if the aspect ratio is large enough and the mass ratio small enough \cite{Tavallaeinejad20}, etc. The method could also be extended to flows subject to a subcritical Hopf or subcritical pitchfork bifurcation. In the latter case, three stable equilibria exist, and some examples are found in the infinitely diverging channel (Jeffery-Hamel flow) \cite{Kerswell04},  in active suspensions \cite{Ohm22}, in an axisymmetric liquid bridge subjected to axial flow \cite{Lowry97} and many others.

The method to derive a stochastically forced Stuart-Landau equation directly from the stochastically forced Navier-Stokes equations is outlined in $\S$~\ref{sec:Wnlexp}. The probability density function of its solution, as well as the statistics of the transition time between its two deterministic attractors, will then be computed by means of the Fokker-Planck equation. The results are reported and compared with direct numerical simulations in $\S$~\ref{sec:Appli}.

%\vspace{-0.6cm}
%%%%%%%%%%%%%%%%%%%%%%%%%%%%%%%%%%%%%%%%%%
%%%%%%%%%%%%%%%%%%%%%%%%%%%%%%%%%%%%%%%%%%
%%%%%%%%%%%%%%%%%%%%%%%%%%%%%%%%%%%%%%%%%%
%%%%%%%%%%%%%%%%%%%%%%%%%%%%%%%%%%%%%%%%%%
\section{Problem definition \label{sec:Pbdef}}
%%%%%%%%%%%%%%%%%%%%%%%%%%%%%%%%%%%%%%%%%%
%%%%%%%%%%%%%%%%%%%%%%%%%%%%%%%%%%%%%%%%%%
%%%%%%%%%%%%%%%%%%%%%%%%%%%%%%%%%%%%%%%%%%
%%%%%%%%%%%%%%%%%%%%%%%%%%%%%%%%%%%%%%%%%%
We consider fluid flows governed by the incompressible Navier-Stokes equations (NSE)
%%%
\begin{equation}
\begin{split}
\pa_t \bU &= -C[\bU,\bU] - \nab P + \Ren^{-1} \Delta \bU + F  \xi(\al \e t) \bff \\
0 &= \nab \cdot \bU,
\label{eq:ns}
\end{split}
\end{equation}
%%%
where $\bU$ is the velocity field, $P$ is the pressure field ensuring $\bU$ to be divergence-free, and $\Ren$ is the Reynolds number. The nonlinear advection operator $C[\ba,\bb] \doteq ((\nab \ba)\bb + (\nab \bb)\ba)/2$ has been defined. We will restrict the analysis to $\Ren$ numbers asymptotically close to a critical value $\Renc$, where the flow experiences a steady and supercritical symmetry-breaking bifurcation, also called supercritical ``pitchfork" bifurcation. Note that in the rest of the paper, the use of the term ``symmetry" will always refer to a discrete symmetry. Specifically, we will consider cases where a distance to criticality, defined as 
%%%
\begin{equation}
\begin{split}
\e \doteq \Renc^{-1} - \Ren^{-1}, \quad \text{is such that } \quad 0<\e \ll 1,
%
%\e \doteq  \Renc^{-1} - \Ren^{-1}  \ \text{is such that } \ \e \ll 1,
%
\label{eq:eps}
\end{split}
\end{equation}
%%%
in accordance with \cite{Sipp07,Camarri19}. In other terms, the symmetric flow at $\Ren>\Renc$ possesses a steady symmetry-breaking eigenmode unstable with a growth rate of $O(\e)$. In the linear regime, this mode thus grows exponentially until nonlinearities have an effect after a long time of $O(\e^{-1})$, implying its amplitude to be both linearly and nonlinearly modulated over a slow time scale $\tau \doteq \e t$. 

In addition, the flow is stochastically forced with $F\xi(\al \e t)\bff$, where $\bff=\bff(\bx)$ is the forcing spatial structure. The term $\xi(\al \e t)$, with $\al>0$ and $\al = O(1)$, is a random signal with zero mean. Since $\e$ is by assumption very small, this signal varies slowly as compared to $\xi(t)$, and according to the same slow time $\tau$ as that of the symmetry-breaking eigenmode. The signal $\xi(t)$ is a Gaussian sampling-limited white noise signal. The latter has a typically large band-limiting frequency $\om_d$, given by the sampling time step $\Delta t$ as $\om_d = \pi/\Delta t$. \newline
%%%%%%%%%%%%
%%%%%%%%%%%%
%%%%%%%%%%%%
%%%%%%%%%%%%
In order to characterize these random processes, we introduce the Fourier transform $\Fo{\ast}$ of a temporal signal of length $[0,T]$ with $T\rightarrow \infty$, and its inverse $\Foi{\ast}$ as
%%%%%%%%%%%%%%%%%%%
%%%%%%%%%%%%%%%%%%%
\begin{equation}
\hbu(\om) = \Fo{\bu(t)} = \frac{1}{\sqrt{T}}\int_{0}^{T}\bu(t)e^{-i\om t}\di t, \quad \bu(t) = \Foi{\hbu(\om)} = \frac{\sqrt{T}}{2\pi}\int_{-\infty}^{\infty}\hbu(\om)e^{i\om t}\di \om.
\label{eq:fft}
\end{equation}
%%%%%%%%%%%%%%%%%%%
%%%%%%%%%%%%%%%%%%%
In the Fourier domain, the random signal $\hxi(\om) = \Fo{\xi(t)}$ is constructed as 
%%%%%%%%%%%%%%%%%%%
%%%%%%%%%%%%%%%%%%%
\begin{equation}
|\hxi(\om)|= \al \sqrt{\e}, \ \text{for}  \  |\om|\leq \om_d, \ \text{and} \ |\hxi(\om)|=0 \ \text{elsewhere}. 
\label{eq:xit}
\end{equation}
%%%%%%%%%%%%%%%%%%%
%%%%%%%%%%%%%%%%%%%
Note that $\al \sqrt{\e}$ does not depend on the frequency. The choice of this specific value for the intensity of $\xi(t)$ is made purposely so that the Fourier transform of the noise $\xi(\al\e t)$, when taken over the slow time $\tau=\e t$, yields a unit intensity. Since $\al$ is not included in the definition of the slow time $\tau$, its square root is not taken in (\ref{eq:xit}), whereas that of $\e$ is. The calculations will be provided in the next section. For each $\om$, the phase of $\hxi(\om)$ is random and drawn according to a uniformly distributed law between $0$ and $2\pi$. 

 We emphasize that the stochastic forcing considered in the present paper is not general but specific for at least two reasons. The first is that it consists of a scalar noise process depending solely on time, multiplying a structure that is frozen in space. The second is that the (sampling-limited) white noise defined in (\ref{eq:xit}) has a constant value for $|\hxi(\om)|$, whereas it would be more generally characterized by a constant value for $\ea{|\hxi(\om)|^2}$, if $\ea{\ast}$ denotes the ensemble average. This generalisation matters in the cases where $T$ is finite and/or if the noise is not an ergodic process.

From the definition (\ref{eq:fft}), it follows that
%%%%%%%%%%%%%%%%%%%
%%%%%%%%%%%%%%%%%%%
\begin{equation}
\begin{split}
\Fo{\xi(\al \e t)} &= \frac{1}{\sqrt{T}}\int_{0}^{T}\xi(\al \e t)e^{-i\om t}\di t = \frac{1}{\al \e \sqrt{T}}\int_{0}^{\al \e T}\xi(s)e^{-is\om/(\al \e)}\di s = \frac{1}{\al \e}\hxi\left(\frac{\om}{\al \e}\right),
\end{split}
\label{eq:xiet}
\end{equation}
%%%%%%%%%%%%%%%%%%%
%%%%%%%%%%%%%%%%%%%
(which is a well-known property of the Fourier transform) therefore,
%%%%%%%%%%%%%%%%%%%
%%%%%%%%%%%%%%%%%%%
\begin{equation}
\begin{split}
|\Fo{\xi(\al \e t)}| = \frac{1}{\sqrt{\e}}  \ \text{for}  \  |\om|\leq \om_{co},
\end{split}
\end{equation}
%%%%%%%%%%%%%%%%%%%
%%%%%%%%%%%%%%%%%%%
where we have defined 
%%%%%%%%%%%%%%%%%%%
%%%%%%%%%%%%%%%%%%%
\begin{equation}
\om_{co} \doteq \al \e \om_d,  \quad \text{and where we recall that} \quad \al=O(1).
\label{eq:Alscale}
\end{equation}
%%%%%%%%%%%%%%%%%%%
%%%%%%%%%%%%%%%%%%%
Therefore, the frequency $\om_{co}$ is the cut-off frequency ( hence the subscript ``co" ) of the slowly varying noise $\xi(\al \e t)$. Although it was implicit, in (\ref{eq:xiet}) we have also used the fact that, since we take the limit of infinitely large $T$ and $\al,\e >0$, integrating between $0$ and $T$ or between $0$ and $\al \e T$ leads to the same result. The small parameter $\e$ being set by the $\Ren$ number, the parameter $\al$ gives the freedom to change $\om_{co}$ as long as $\al$ is of order unity. Accordingly, the small parameter $\al \e = \om_{co}/\om_d = O(\e) \ll 1$ takes the immediate meaning of the ratio between the shortest measurable time scale (i.e., the sampling one $=\pi/\om_d$), and the shortest time scale excited by $\xi(\al \e t)$.

%%%%%%%%%%%%%%%%%%%
%%%%%%%%%%%%%%%%%%%
\begin{figure} 
\centering
  \begin{subfigure}[b]{0.44\linewidth}
\includegraphics[trim={0cm 0.cm 0cm 0cm},clip,width=1\linewidth]{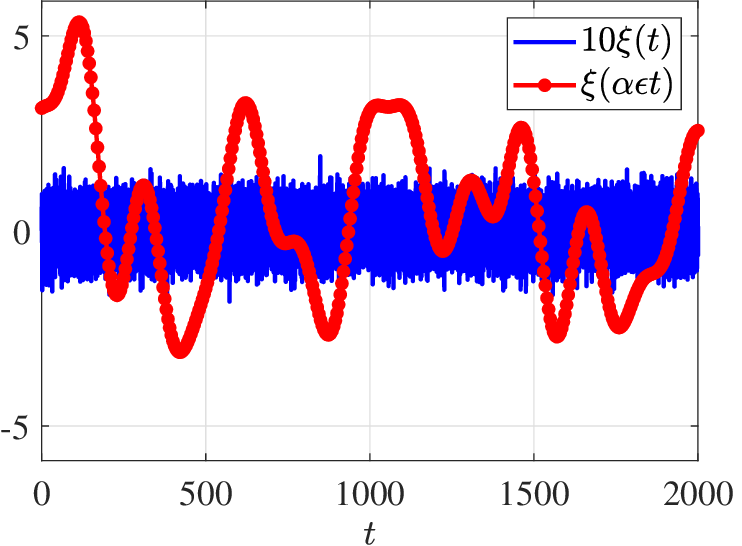}% Here is how to import EPS art
\caption{\label{fig:1a}}
\end{subfigure}
%%%%%%%%%%%%%%%%%%%
  \hspace{1cm}
%%%%%%%%%%%%%%%%%%%  
  \begin{subfigure}[b]{0.445\linewidth}
\includegraphics[trim={0cm 0cm 0cm 0cm},clip,width=1\linewidth]{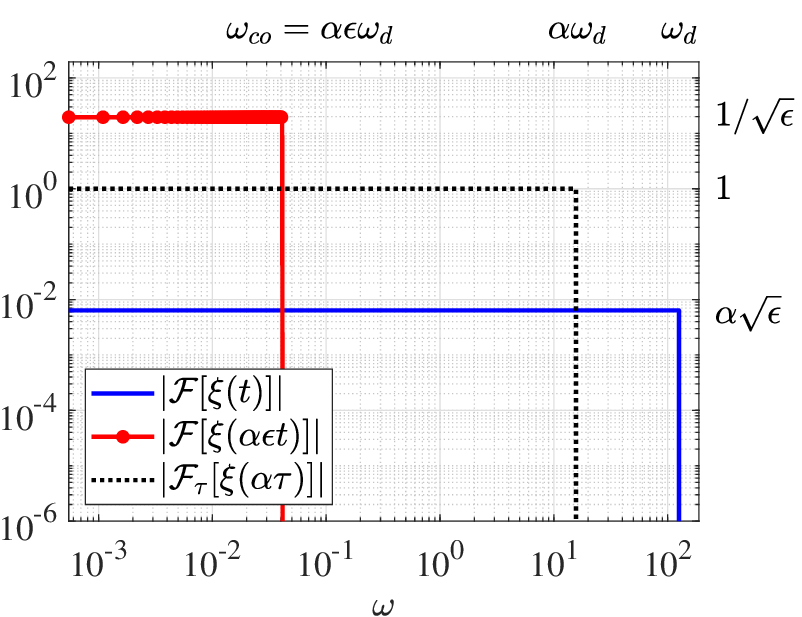}% Here is how to import EPS art
\caption{\label{fig:1b}}
 \end{subfigure}
%%%%%%%%%%%%%%%%%%% 
\caption{(a) Example of random signals as a function of time. The blue continuous line is a sampling-limited white noise $\xi(t)$ (multiplied by a factor $10$ for better visualization), whereas the blue continuous line with bullets is the slowly varying version $\xi(\al \e t)$. (b) Module of the Fourier transforms of the signals shown in (a). The specific values $\e=0.0026$, $\alpha=1/8$ and a sampling time step $\Delta t = 0.025$ have been selected; this sets $\om_d = 125.7$ and $\om_{co}=0.04$.\label{fig:1} }
\end{figure}
%%%%%%%%%%%%%%%%%%%
%%%%%%%%%%%%%%%%%%% 

 Some random signals $\xi(t)$ and $\xi(\al \e t)$ are shown in figure~\ref{fig:1} in both the temporal and Fourier domains. We stress that, as long as the slow noise is band-limited with a cut-off frequency around $\om_{co}$, the results are not expected to depend on the specific shape of its spectrum (see also a more detailed discussion in section IV.D.)

We also introduce the Hermitian inner products
%%%%%%%%%%%%%%%%%%%
%%%%%%%%%%%%%%%%%%%
\begin{equation}
 \ssp{\bu_a}{\bu_b} \doteq \int_{\Omega} \bu_a^{\Her}\bu_b\di \Omega \ \ \text{and} \ \ \ssp{\mat{\bu_a}{p_a}}{\mat{\bu_b}{p_b}}_{p} \doteq \int_{\Omega} \bu_a^{\Her}\bu_b + p_a^{\Her}p_b\di \Omega,
\label{eq:ip}
\end{equation}
%%%%%%%%%%%%%%%%%%%
%%%%%%%%%%%%%%%%%%%
where the superscript $\Her$ denotes the Hermitian transpose and the symbol $\Omega$ the spatial domain. In (\ref{eq:ip}) the second inner product includes the pressure (hence the subscript $p$) whereas the first doesn't. In the following, $||\ast||$ designates the norm induced by the first scalar product in (\ref{eq:ip}). The spatial structure of the forcing in velocity $\bff$ has a unitary norm, i.e. $||\bff||=1$, and we scale the forcing amplitude $F$ as 
%%%%%%%%%%%%%%%%%%%
%%%%%%%%%%%%%%%%%%%
\begin{equation}
F\doteq \phi \sqrt{\e}^3 \ll 1 \quad \text{with} \quad \phi=O(1).
\label{eq:Fscale}
\end{equation}
%%%%%%%%%%%%%%%%%%%
%%%%%%%%%%%%%%%%%%%

The characterization of the parameters $\e$, $\al$, and $\phi$ is summarized in table~\ref{tab:table1}. Their respective presence gives sufficient freedom for the $\Ren$ number, the slow noise cut-off frequency $\om_{co}$, and the forcing amplitude $F$, to be chosen independently of each other. For instance, a noise with the same intensity and cut-off frequency could force flows with two different $\Ren$ numbers; in the latter case, the parameter $\e$ would be different between the two flows but $\al$ and $\phi$ could be adapted to conserve $\om_{co}$ and $F$.
%%%%%%%%%%%%%%%%%
%%%%%%%%%%%%%%%%%
%%%%%%%%%%%%%%%%%
\begin{table}[tb]%The best place to locate the table environment is directly after its first reference in text
\caption{\label{tab:table1}%
Characterization of the dimensionless parameters}.\begin{ruledtabular}
\begin{tabular}{ccc}
\textrm{Parameter}&
\textrm{Mathematical constraint}&
\textrm{Permits to fix independently ...}\\
\colrule
$\e$ & $\ll 1$ & ... the $\Ren$ number, according to (\ref{eq:eps}) \\
$\al$ & $O(1)$ & ... the noise cut-off frequency $\om_{co}$, according to (\ref{eq:Alscale}) \\
$\phi$ & $O(1)$ & ... the stochastic forcing amplitude $F$, according to (\ref{eq:Fscale})\\
\end{tabular}
\end{ruledtabular}
\end{table}
%%%%%%%%%%%%%%%%%
%%%%%%%%%%%%%%%%%
%%%%%%%%%%%%%%%%%

%%%%%%%%%%%%%%%%%%%%%%%%%%%%%%%%%%%%%%%%%%
%%%%%%%%%%%%%%%%%%%%%%%%%%%%%%%%%%%%%%%%%%
%%%%%%%%%%%%%%%%%%%%%%%%%%%%%%%%%%%%%%%%%%
%%%%%%%%%%%%%%%%%%%%%%%%%%%%%%%%%%%%%%%%%%
\section{Weakly nonlinear expansion \label{sec:Wnlexp}}
%%%%%%%%%%%%%%%%%%%%%%%%%%%%%%%%%%%%%%%%%%
%%%%%%%%%%%%%%%%%%%%%%%%%%%%%%%%%%%%%%%%%%
%%%%%%%%%%%%%%%%%%%%%%%%%%%%%%%%%%%%%%%%%%
%%%%%%%%%%%%%%%%%%%%%%%%%%%%%%%%%%%%%%%%%%
We now derive an equation for the amplitude of the bifurcated steady mode, in the presence of a weak and slowly-varying noise. In the absence of this stochastic forcing, the calculations would be similar in all respects to those of \cite{Camarri19}. Moreover, the derivation of an amplitude equation for the Duffing oscillator subject to a narrow-band noise was proposed already in \cite{Rajan88, Nayfeh90, Rong98}, among others. The rigorous procedure outlined in these latter works to include a noise term in an amplitude equation is applied to the Navier-Stokes equation thereafter.

As already mentioned, since we consider $\Ren$ numbers close to a critical value for a steady bifurcation, as expressed in (\ref{eq:eps}), the temporal variations of the flow perturbation around the neutral equilibrium are assumed to occur over the slow time scale $\tau = \e t$. This assumption and the ensuing scaling are consistent with the fact that the flow is forced by the slow noise $\xi(\al \e t)=\xi(\al \tau)$. The flow field is approached by the following expansions
%%%%%%%%%%%%%%%%%%%
%%%%%%%%%%%%%%%%%%%
\begin{equation}
\begin{split}
\bU &=  \bU_{\tc} + \sqrt{\e} \bu_1(\tau) + \e \bu_2(\tau) + \sqrt{\e}^3\bu_3(\tau) + O(\e^2), \ \text{and} \\
P &=  P_{\tc} + \sqrt{\e} p_1(\tau) + \e p_2(\tau) + \sqrt{\e}^3 p_3(\tau) + O(\e^2),
\label{eq:as}
\end{split}
\end{equation}
%%%%%%%%%%%%%%%%%%%
%%%%%%%%%%%%%%%%%%%
where $\bU_{\tc}(\bx)$ is a velocity field in a neutral equilibrium at $\Renc$, symmetric in space around a generic plane for three-dimensional flows, or around a generic axis for two-dimensional flows. Considering a two-dimensional flow with a symmetry axis located at $y= \ys$, the symmetry assumption of $\bU_{\tc}$ implies
%%%%%%%%%%%%%%%%%%%
%%%%%%%%%%%%%%%%%%%
\begin{equation}
\begin{split}
\bU_{\tc}(\bx) = \mat{U_{\tc}(x,y)}{V_{\tc}(x,y)} = \mat{U_{\tc}(x,2\ys-y)}{-V_{\tc}(x,2\ys-y)}.
\label{eq:sym}
\end{split}
\end{equation}
%%%%%%%%%%%%%%%%%%%
%%%%%%%%%%%%%%%%%%%
By increasing the $\Ren$ number above $\Renc$, the flow $\bU_{\tc}$ is subject to a steady bifurcation which breaks the symmetry of the overall flow.

Introducing the expansions (\ref{eq:eps}) and (\ref{eq:as}) into (\ref{eq:ns}) leads to a cascade of linear problems to be solved successively. At order $O(\sqrt{\e})$, we collect
%%%%%%%%%%%%%%%%%%%
%%%%%%%%%%%%%%%%%%%
\begin{equation}
\begin{split}
\mat{{\bf 0}}{0} = 
L \mat{\bu_1}{p_1} \ \text{with} \ L \doteq \begin{bmatrix}
-2C[\bU_{\tc},\ast] + \Renc^{-1} \Delta & -\nab \\ 
\nab \cdot & 0
\end{bmatrix},  \ \text{therefore}  \ \mat{\bu_1(\tau,\bx)}{p_1(\tau,\bx)} = A(\tau)\mat{\bq(\bx)}{p_{\tq}(\bx)},
\end{split}
\label{eq:o1}
\end{equation}
%%%%%%%%%%%%%%%%%%%
%%%%%%%%%%%%%%%%%%%
where $[\bq(\bx),p_{\tq}]^T$ is the eigenmode of the linear operator $L$ that is associated with a null eigenvalue. We normalize it such that $||\bq||=1$. Note that $\bq$ is also the non-trivial kernel of $L$. In addition, the eigenmode $\bq$ is anti-symmetric: therefore, again considering a two-dimensional flow and a symmetry axis at $y =\ys$, the velocity field $\bq$ satisfies
%%%%%%%%%%%%%%%%%%%
%%%%%%%%%%%%%%%%%%%
\begin{equation}
\begin{split}
\bq(\bx) = \mat{q_x(x,y)}{q_y(x,y)} = \mat{-q_x(x,2\ys-y)}{q_y(x,2\ys-y)}.
\end{split}
\label{eq:antisym}
\end{equation}
%%%%%%%%%%%%%%%%%%%
%%%%%%%%%%%%%%%%%%%
In the following, the adjoint mode associated with $[\bq,p_{\tq}]^T$ will be denoted by $[\bq^{\da},p^{\da}_{\tq}]^T$. In other terms, it corresponds to the eigenmode associated with the null eigenvalue of the operator $L^{\da}$, adjoint to $L$ under the second scalar product in (\ref{eq:ip}). The slowly varying and real scalar amplitude $A(\tau)$ in (\ref{eq:o1}) is for now arbitrary.

At $O(\e)$ we obtain the solution $\bu_2(\tau,\bx) = A(\tau)^2\buR(\bx) + \buV(\bx)$, where
%%%%%%%%%%%%%%%%%%%
%%%%%%%%%%%%%%%%%%%
\begin{equation}
\begin{split}
-L\mat{\buR}{\pR} = \mat{-C[\bq,\bq]}{0}, \ \text{and} \ -L\mat{\buV}{\pV} = \mat{-\Delta \bU_{\tc}}{0}.
\end{split}
\label{eq:o2}
\end{equation}
%%%%%%%%%%%%%%%%%%%
%%%%%%%%%%%%%%%%%%%
The operator $L$ being singular, the compatibility condition needs to be verified for the particular solutions to the systems in (\ref{eq:o2}) to be non-diverging. The latter condition requires the right-hand side to be orthogonal to the kernel of the adjoint of $L$, i.e. orthogonal to $[\bq^{\da},p_{\tq}^{\da}]$. The symmetric fields $C[\bq,\bq]$ and $\Delta \bU_{\tc}$ yielding a null inner product with the anti-symmetric one $\bq^{\da}$, this condition is naturally satisfied and the two systems can directly be solved for. In practice, this can for instance be done with a pseudo-inverse algorithm. The component of $\buR$ and $\buV$ on the kernel $\bq$ that stems from the homogeneous part of the solution is set to zero according to $\ssp{\bq^{\da}}{\buR}=\ssp{\bq^{\da}}{\buV}=0$. Indeed, accounting for a non-zero homogeneous solution was shown in \cite{Camarri19} to have no consequences on the coefficients of the final amplitude equation.

At $O(\sqrt{\e}^3)$ we assemble the system
%%%%%%%%%%%%%%%%%%%
%%%%%%%%%%%%%%%%%%%
\begin{equation}
\begin{split}
-L\mat{\bu_3}{p_3} = -A\mat{2C[\bq,\buV] + \Delta  \bq }{0} - A^3\mat{2C[\bq,\buR]}{0} - \frac{\di A}{\di \tau} \mat{\bq}{0} + \phi \xi(\al \tau)\mat{\bff}{0}.
\end{split}
\label{eq:o3}
\end{equation}
%%%%%%%%%%%%%%%%%%%
%%%%%%%%%%%%%%%%%%%
This time, the compatibility condition is not naturally satisfied but leads to an equation for $A(\tau)$
%%%%%%%%%%%%%%%%%%%
%%%%%%%%%%%%%%%%%%%
\begin{equation}
\begin{split}
\frac{\di A}{\di \tau} = \lambda A(\tau) + \mu A(\tau)^3 + \eta \phi \xi(\al \tau) = -\frac{\di V}{\di A} + \eta \phi \xi(\al \tau) ,
\end{split}
\label{eq:ampeq}
\end{equation}
%%%%%%%%%%%%%%%%%%%
%%%%%%%%%%%%%%%%%%%
with the coefficients
%%%%%%%%%%%%%%%%%%%
%%%%%%%%%%%%%%%%%%%
\begin{equation}
\begin{split}
\lambda = -\frac{\ssp{\bq^{\da}}{2C[\bq,\buV] + \Delta  \bq}}{\ssp{\bq^{\da}}{\bq}}, \quad \mu = -\frac{\ssp{\bq^{\da}}{2C[\bq,\buR]}}{\ssp{\bq^{\da}}{\bq}} \quad \text{and} \quad \eta = \frac{\ssp{\bq^{\da}}{\bff}}{\ssp{\bq^{\da}}{\bq}}.
\end{split}
\label{eq:coeffs}
\end{equation}
%%%%%%%%%%%%%%%%%%%
%%%%%%%%%%%%%%%%%%%
In addition, the double-well potential 
%%%%%%%%%%%%%%%%%%%
%%%%%%%%%%%%%%%%%%%
\begin{equation}
\begin{split}
V = V[A] \doteq -\frac{\lambda A^2}{2} -  \frac{\mu A^4}{4} 
\end{split}
\label{eq:pot}
\end{equation}
%%%%%%%%%%%%%%%%%%%
%%%%%%%%%%%%%%%%%%%
has been defined. Equation (\ref{eq:ampeq}), for the amplitude of the anti-symmetric bifurcated mode, is the classic Stuart-Landau equation for a real-valued amplitude, with the difference that it is now stochastically forced. The stochastic term was not an \textit{ad hoc} addition to a pre-existing amplitude equation, but was derived rigorously from the forced Navier-Stokes equations (\ref{eq:ns}). The explicit formulas for the coefficients in (\ref{eq:coeffs}) can be directly evaluated numerically and do not require any \textit{a posteriori} fitting. Indeed, they involve scalar products of fields that are all known: the eigenmode $\bq$, the adjoint mode $\bq^{\da}$, the second-order fields $\buV$ and $\buR$, all defined at $\Ren=\Renc$, as well as the forcing structure $\bff$.

The Fourier transform over the slow time scale, noted $\Fos{\tau}{\ast}$, of the noise $\xi(\al \tau)$ reads 
%%%%%%%%%%%%%%%%%%%
%%%%%%%%%%%%%%%%%%%
\begin{equation}
\begin{split}
\Fos{\tau}{\xi(\al \tau)} = \frac{1}{\sqrt{\e T}}\int_{0}^{\e T}\xi(\al \tau) e^{-i\om \tau}\di \tau = \frac{\sqrt{\e}}{\sqrt{ T}}\int_{0}^{ T}\xi(\al \e t) e^{-i\om \e t}\di t = \frac{1}{\al \sqrt{\e} \sqrt{ T}}\int_{0}^{\al \e T}\xi(s) e^{-i\om s/\al}\di s = \frac{1}{\al \sqrt{\e}}\hxi\left(\frac{\om}{\al} \right),
\end{split}
\label{eq:Fos}
\end{equation}
%%%%%%%%%%%%%%%%%%%
%%%%%%%%%%%%%%%%%%%
therefore, following (\ref{eq:xit}), we have that 
%%%%%%%%%%%%%%%%%%%
%%%%%%%%%%%%%%%%%%%
\begin{equation}
\begin{split}
|\Fos{\tau}{\xi(\al \tau)}| = 1 \quad \text{for} \quad |\om| \leq \al \om_d,
\end{split}
\label{eq:altau}
\end{equation}
%%%%%%%%%%%%%%%%%%%
%%%%%%%%%%%%%%%%%%%
and $|\Fos{\tau}{\xi(\al \tau)}| = 0$ everywhere else. This profile is illustrated by the black dotted line in figure~\ref{fig:1b}. It is important to notice that $\Fos{\tau}{\xi(\al \tau)}$ is independent of the small parameter $\e$, which was done intentionally and explains the peculiar choice of intensity in (\ref{eq:xit}). In this manner, the amplitude equation (\ref{eq:ampeq}) does not depend on $\e$ and the ensuing results need not be re-computed for each $\Ren$ considered, if everything else is fixed.

Let us briefly discuss the deterministic regime where $\phi=0$. The coefficient $\lambda$ contains the sensitivity of the null eigenvalue to a base flow modification $+ \e \buV$ induced by the fact that we consider a $\Ren > \Renc$. Therefore, $\e \lambda$ is directly the growth rate of the bifurcated steady mode and is positive. Accordingly, the equilibrium solution $\bar{A}_0=0$ of (\ref{eq:ampeq}), corresponding to a symmetric flow, is unstable. The coefficient $\mu$ contains the sensitivity of the null eigenvalue to a base flow modification $+\e A^2 \buR$, non-linearly induced by the Reynolds stress of the perturbation $\sqrt{\e}A\bq$. If $\mu$ is negative, then nonlinearities have a stabilizing effect that counteracts linear instability. Therefore two additional equilibrium solutions $\pm \bar{A}$, with $\bar{A}\doteq \sqrt{-\lambda/\mu}>0$, exist and are stable. They are the two minima of the potential $V$. On the contrary, if $\mu$ is positive, nonlinearities included in (\ref{eq:ampeq}) have a destabilizing effect and no stable equilibrium exists at that order: the bifurcation is subcritical and the expansion must be pursued at higher orders. This latter case will not be treated in what follows.

In the stochastically forced regime where $\phi \neq 0$, the amplitude $A$ may randomly switch back and forth between the two attractors $\bar{A}$ and $-\bar{A}$ after unpredictable and possibly long times. Therefore, we are particularly interested in the probability distribution of $A$. The probability density function (PDF) $P$ of the amplitude $A$ can be computed directly by means of the Fokker-Planck equation \cite{Risken96}. For this, the state space needs to be augmented in order to account for the fact that $\xi(\al \tau)$ is a band-limited white noise because of the presence of the constant $\al$ in the argument. Following \cite{Risken96} (Appendix A1 and Supplement S.10), $\xi(\al \tau)$ is treated as a system variable resulting from low-pass filtering, at a cut-off frequency of $\al \om_d$, of a white noise (on the slow time scale) $\chi(\tau)$ whose band-limiting frequency is the largest achievable by the system, $\om_d$. We insist that $\om_d$ is chosen sufficiently large for none of the results presented in this paper to depend on it. For the sake of simplicity, a first-order low-pass filter of cut-off frequency $\al \om_d$ is chosen and thus $\xi(\al \tau)$ is approached by the solution of the first order stochastic differential equation 
%%%%%%%%%%%%%%%%%%%
%%%%%%%%%%%%%%%%%%%
\begin{equation}
\begin{split}
\frac{\di \xi}{\di \tau} + \al \om_d \xi = \al \om_d \chi,
\end{split}
\label{eq:lwp}
\end{equation}
%%%%%%%%%%%%%%%%%%%
%%%%%%%%%%%%%%%%%%%
where $\chi(\tau)$ is a white noise such that $|\Fos{\tau}{\chi}| = 1$ for $|\om| \leq \om_d$. Equation (\ref{eq:lwp}) leads to $|\Fos{\tau}{\xi}|^2=1/(1+\om^2/(\al \om_d)^2)$, which is a rather coarse approximation of (\ref{eq:altau}). This matters little, however, since as mentioned, the results aren't expected to depend too much on the exact shape of the noise as long as they conserve the cut-off frequency. Again following \cite{Risken96} (Chapter 4.7), the Fokker-Planck equation associated with the system consisting of the equations (\ref{eq:ampeq}) and (\ref{eq:lwp}) writes
%%%%%%%%%%%%%%%%%%%
%%%%%%%%%%%%%%%%%%%
\begin{equation}
\begin{split}
\frac{\pa P}{\pa \tau} = -\frac{\pa}{\pa A}\left[(\lambda A + \mu A^3 + \eta \phi \xi ) P \right] + \al \om_d \frac{\pa \left( \xi P \right)}{\pa \xi} + \frac{(\al \om_d)^2}{2}\frac{\pa^2 P}{\pa \xi^2},
\end{split}
\label{eq:FP}
\end{equation}
%%%%%%%%%%%%%%%%%%%
%%%%%%%%%%%%%%%%%%%
where $P=P[A,\xi,\tau]$ vanishes for $|A|, |\xi| \rightarrow \infty$. By definition of a probability density function, $P$ has a unitary area: $\iint_{-\infty}^{\infty} P \di A \di \xi = 1$, $\forall \tau$. We introduce $\PS = \PS[A]$ the PDF of only $A$ in a stationary regime, reached after infinitely long $\tau$, such that
%%%%%%%%%%%%%%%%%%%
%%%%%%%%%%%%%%%%%%%
\begin{equation}
\begin{split}
\PS \doteq \int_{-\infty}^{\infty} \left(\lim_{\tau \rightarrow \infty}P[A,\xi,\tau]\right) \di \xi. 
\label{eq:ps}
\end{split}
\end{equation}
%%%%%%%%%%%%%%%%%%%
%%%%%%%%%%%%%%%%%%%
We also define $\PSw$ to be $\PS$ in the ``pure" white noise limit where $\al \om_d \rightarrow \infty$, which possesses the analytical expression
%%%%%%%%%%%%%%%%%%%
%%%%%%%%%%%%%%%%%%%
\begin{equation}
\begin{split}
\PSw \doteq \lim_{\al \om_d \rightarrow \infty} \PS = \frac{1}{Z} \exp\left(-\frac{2V}{(\phi\eta)^2} \right), \quad \text{$Z$ a normalization constant.} 
\label{eq:PSw}
\end{split}
\end{equation}
%%%%%%%%%%%%%%%%%%%
%%%%%%%%%%%%%%%%%%%
In the next section, the stochastically forced weakly nonlinear (WNL) amplitude $A$, its probability density function $\PS$, and the statistics of the escape time $\Delta T$ of $A$ between its two attractors $\pm\bar{A}$ will be computed for a selected flow geometry. Furthermore, results will be compared to direct numerical simulations (DNS). 
%%%%%%%%%%%%%%%%%%%
%%%%%%%%%%%%%%%%%%%
%%%%%%%%%%%%%%%%%%%
\section{Application case: the flow past a sudden expansion \label{sec:Appli}}
%%%%%%%%%%%%%%%%%%%
%%%%%%%%%%%%%%%%%%%
%%%%%%%%%%%%%%%%%%%
The application case is chosen as the two-dimensional plane flow past a sudden expansion (see the non-dimensional geometry in figure~\ref{fig:2a}). The Reynolds number is defined as $\Ren = h U_{\infty}/\nu$, where $h$ is the inlet channel height, $U_{\infty}$ the centreline (maximum) velocity of the inlet parabolic velocity profile and $\nu$ the kinematic viscosity. The inlet is located at the streamwise coordinate $x=-5$ (made non-dimensional by $h$). At $x=0$, the flow goes through a sudden expansion of expansion ratio $\text{ER}=3$, and the outlet of the expansion is situated further downstream at $x=L=40$, where the flow re-parallelized.

The linear and nonlinear Navier–Stokes equations are solved for the velocity field $\bU=[U,V]^T$ and the pressure by means of the finite element method with Taylor–Hood (P2, P2, P1) elements, respectively, after implementation of the weak form in the software FreeFem++. The steady solutions of the Navier–Stokes equations are solved using the iterative Newton–Raphson method, and the linear operators are built thanks to a sparse solver UMFPACK implemented in FreeFem++. The mesh is constituted of approximately $4 \times 10^4$ triangular elements, whose edge size varies between a minimum value of $0.015$ near the expansion corners and a maximal value of $0.15$ farther upstream and downstream, leading to about $2 \times 10^5$ degrees of freedom for the global flow field (velocity and pressure). 

%%%%%%%%%%%%%%%%%%%
%%%%%%%%%%%%%%%%%%%
%%%%%%%%%%%%%%%%%%%
\subsection{Deterministic regime}
%%%%%%%%%%%%%%%%%%%
%%%%%%%%%%%%%%%%%%%
%%%%%%%%%%%%%%%%%%%

\begin{figure} 
\centering
  \begin{subfigure}[b]{0.44\linewidth}
\includegraphics[trim={0cm 0.65cm 0cm 0.15cm},clip,width=1\linewidth]{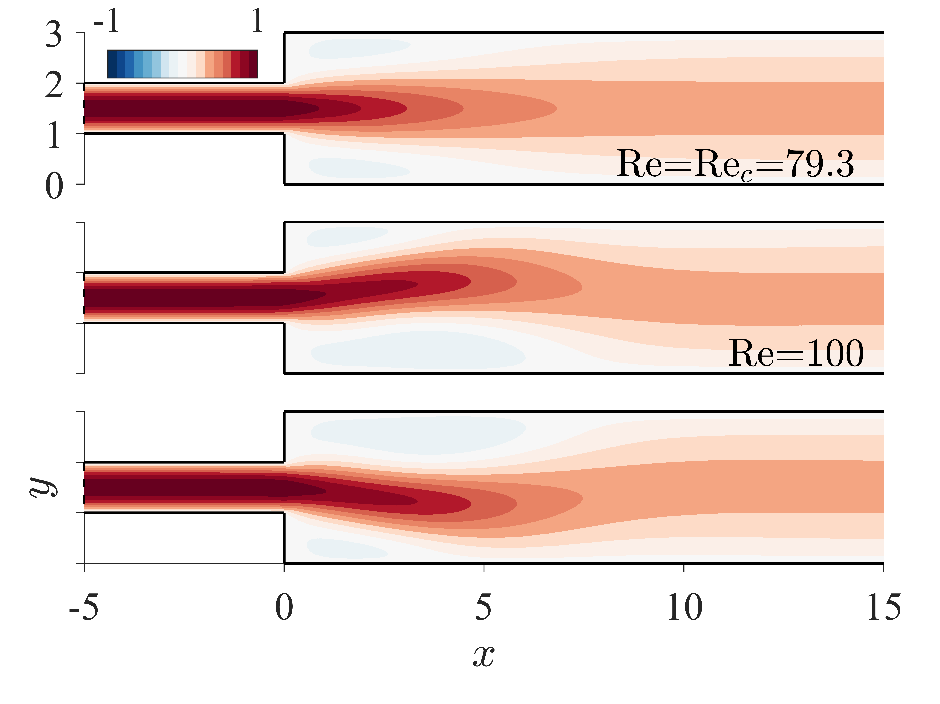}% Here is how to import EPS art
\caption{\label{fig:2a}}
\end{subfigure}
%%%%%%%%%%%%%%%%%%%
  \hspace{1cm}
%%%%%%%%%%%%%%%%%%%  
  \begin{subfigure}[b]{0.44\linewidth}
\includegraphics[trim={0cm 0.0cm 0cm 0.0cm},clip,width=1\linewidth]{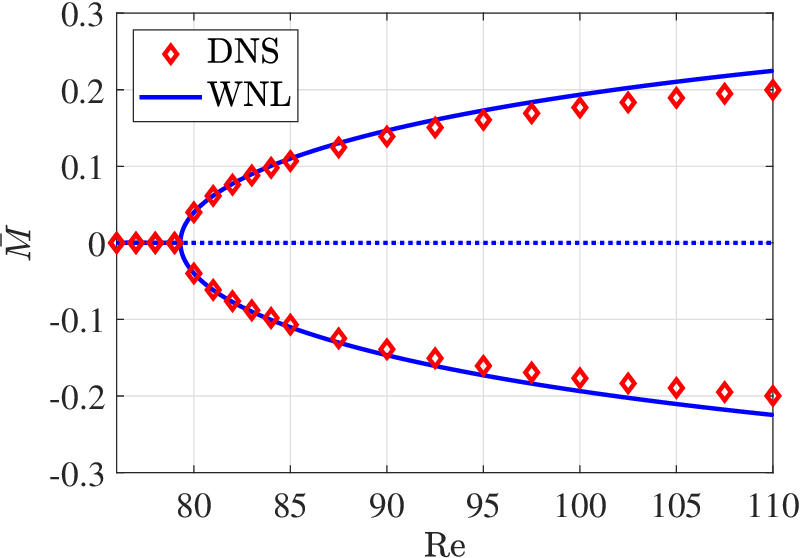}% Here is how to import EPS art
\caption{\label{fig:2b}}
 \end{subfigure}
%%%%%%%%%%%%%%%%%%% 
\caption{Deterministic case $\phi=0$. (a) Snapshots of the streamwise velocity of the stable steady solution(s) obtained by DNS at $\Ren=\Renc=79.3$ (top) and at $\Ren=100>79.3$ (center $\&$ bottom); in the latter case, two equilibrium solutions are found. (b) Measure $M$ that quantifies the asymmetry of the flow, as defined in (\ref{eq:M}), and computed in the steady regime for both WNL and DNS approaches. The continuous blue line is associated with the stable equilibrium solutions predicted by the deterministic amplitude equation $\pm \bar{M}$ with $\bar{M}=\sqrt{\e}\beta \bar{A}$, whereas the dotted line corresponds to the unstable equilibrium. \label{fig:2} }
\end{figure}

The critical $\Ren$ number before the steady bifurcation is numerically found to be $\Renc = 79.3$, which compares relatively well with the value $\Renc = 81.4$ reported in \cite{Camarri19}. As we checked the convergence of $\Renc$ with respect to the spatial discretization, the slight difference is rather explained by the fact that our entrance length is half of the one considered in \cite{Camarri19}. For $\Ren \leq \Renc$, the flow is symmetric in the sense of (\ref{eq:sym}) around the centerline axis at $y=\ys=1.5$, as can be seen in the uppermost snapshot in figure~\ref{fig:2a}. 
%%%%%%%%%%%%%%%%%%%
%%%%%%%%%%%%%%%%%%%
%%%%%%%%%%%%%%%%%%%
%%%%%%%%%%%%%%%%%%%
%%%%%%%%%%%%%%%%%%%
%%%%%%%%%%%%%%%%%%%

For $\Ren > \Renc$, the symmetry of the flow is broken, as we can observe on the snapshots at the center and the bottom of figure~\ref{fig:2a}. The degree of asymmetry is quantified according to a scalar measure that we call $M$, and whose definition is always arguably arbitrary. Nevertheless, we propose it to be the signed $L^2$-norm of the cross-wise velocity component evaluated along the symmetry axis (located at $y=\ys=1.5$). The sign is chosen as being that of the cross-wise velocity along the symmetry axis and at $x=2$, for it is where the cross-wise velocity component $q_y$ of the anti-symmetric eigenmode $\bq=[q_x,q_y]^T$ (with $||\bq||=1$) reaches its (chosen positive) maximum. In other terms, $q_y(x=2,\ys) = \max_{x}[q_y(x,\ys)]=0.136$. Eventually, $M$ reads
%%%%%%%%%%%%%%%%%%%
%%%%%%%%%%%%%%%%%%%
\begin{equation}
\begin{split}
M \doteq \sgn[V(x=2,\ys)]\sqrt{\int_{0}^{L}V(x,\ys)^2\di x}.
\end{split}
\label{eq:M}
\end{equation}
%%%%%%%%%%%%%%%%%%%
%%%%%%%%%%%%%%%%%%%
The WNL approximation of $M$ is given by evaluating $V(x,\ys)$ according to the expansion (\ref{eq:as}), then using that $V_{\tc}(x,\ys)=v_2(x,\ys)=0$, $\forall x$ ($V_{\tc}$ the cross-wise velocity of the base flow and $v_2$ that of the second order field $\bu_2$) for a symmetry reason, leading to 
%%%%%%%%%%%%%%%%%%%
%%%%%%%%%%%%%%%%%%%
\begin{equation}
\begin{split}
M &= \sgn[\underbrace{\sqrt{\e}q_y(2,\ys)}_{>0}A+O(\sqrt{\e}^3)]\sqrt{\int_{0}^{L}\left(\sqrt{\e}A q_y(x,\ys)+O(\sqrt{\e}^3)\right)^2\di x}\\
&= \sgn(A)\sqrt{\e}|A|\sqrt{\int_{0}^{L}q_y(x,\ys)^2\di x} + O(\sqrt{\e}^3)\\
&= \sqrt{\e}\beta A+ O(\sqrt{\e}^3),
\end{split}
\label{eq:Mwnl}
\end{equation}
%%%%%%%%%%%%%%%%%%%
%%%%%%%%%%%%%%%%%%%

with $\beta = \sqrt{\int_{0}^{L}q_y(x,\ys)^2\di x} = 0.266 $ a proportionality constant.

The coefficients in (\ref{eq:coeffs}) are found as $\lambda = 5.984$ and $\mu = -0.02962$, leading to $\bar{A} = \sqrt{-\lambda/\mu}= 14.21$. We stress that the coefficients are evaluated from known fields, without any fitting parameters. For a given $\Ren$ number (setting $\e$), the equilibrium amplitudes $\pm\bar{A}$ are associated with the equilibrium asymmetry measures $\pm\bar{M}$ with $\bar{M}=\sqrt{\e}\beta \bar{A}$. The slope of the red dashed line in figure~4b of \cite{Camarri19} corresponds to our definition of $\lambda$. By visual inspection, we estimate for the former a value of $\approx 5.8$ which indeed agrees well with our $\lambda$; no numerical value is given for $\mu$ in \cite{Camarri19}.

In the deterministic case where $\phi=0$, the DNS and WNL steady solutions are compared under the measure $M$ in figure~\ref{fig:2b} as a function of the $\Ren$ number. Close the threshold value $\Renc$, both approaches are in excellent agreement, thus validating the well-posedness of the weakly nonlinear expansion (\ref{eq:as}). Note that, interestingly, the latter implies the scaling $M \propto \sqrt{1/\Renc-1/\Ren}$ when $\Ren$ is asymptotically close to $\Renc$. The agreement between both approaches progressively degrades as we increase $\Ren$, presumably due to higher-order nonlinearities neglected in the expansion. Nevertheless, the relative error remains reasonable for the considered range of $\Ren$, with a maximum value of $\approx 11 \%$ for $\Ren=110$. Overall, the agreement between both approaches is comparable with the one already reported (with a different measure) in figure $5$ of \cite{Hawa01}. 

In the rest of the study, we will fix the $\Ren$ number to $\Ren=100$.

%%%%%%%%%%%%%%%%%%%
%%%%%%%%%%%%%%%%%%%
%%%%%%%%%%%%%%%%%%%
\subsection{Stochastic regime: amplitude statistics}
%%%%%%%%%%%%%%%%%%%
%%%%%%%%%%%%%%%%%%%
%%%%%%%%%%%%%%%%%%%

\begin{figure}
\centering
\includegraphics[trim={0.0cm 0.0cm 0.0cm 0.0cm},clip,width=0.97\linewidth]{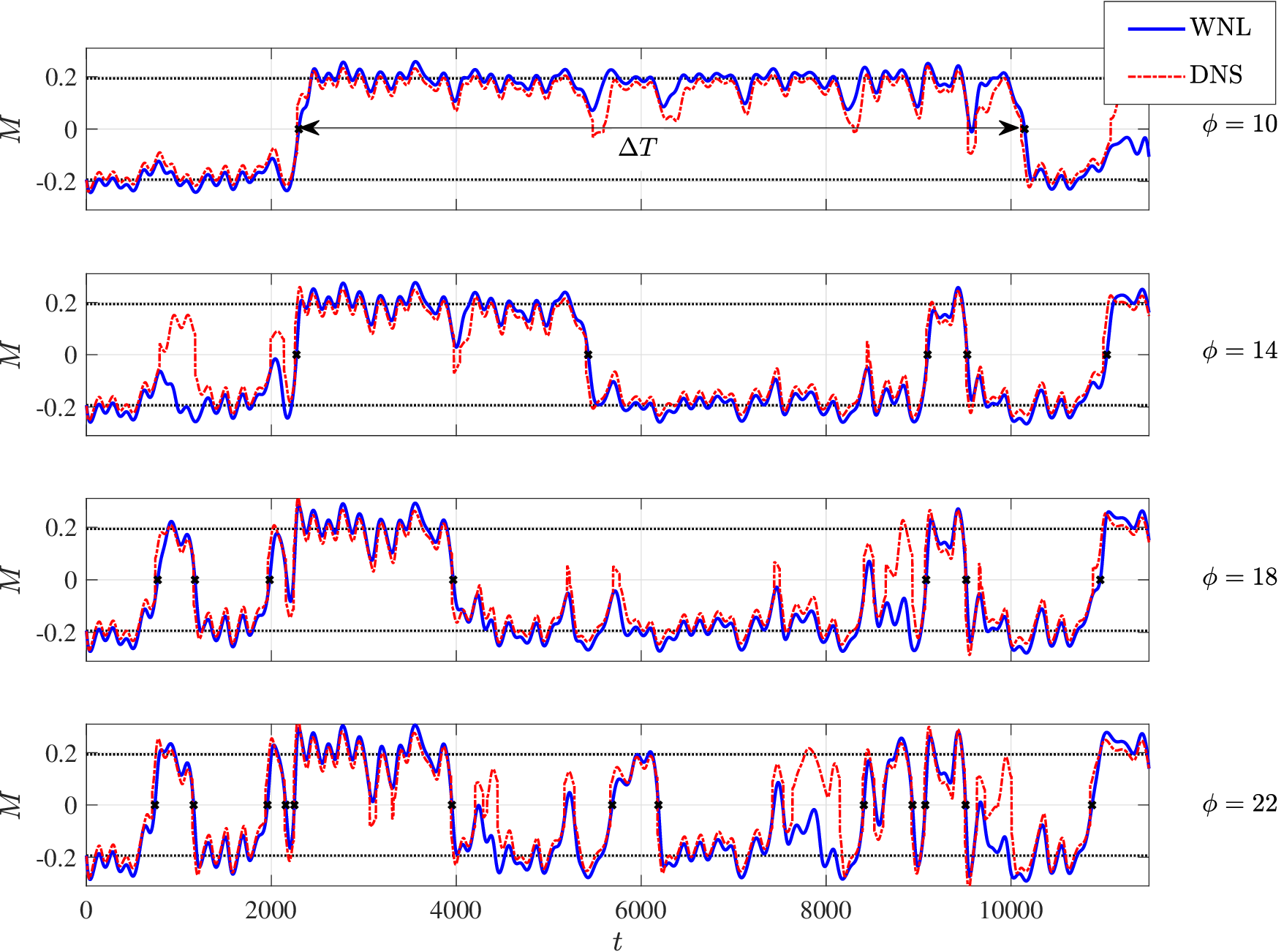}
\caption{Temporal signal of $M$ in a stochastically forced regime, with forcing amplitudes $\phi \in [10,14,18,22]$. The noise realization $\xi(\al \e t)$ is the same for all the results shown. A $\Ren$ number of $\Ren=100$ is chosen, implying $\e=0.0026$. A time step $\Delta t = 0.025$ was found sufficient for the convergence of the results, which sets $\om_d = 125.7$. We further choose $\om_{co}=0.04$, which determines $\alpha=1/8$ that we check to be $O(1)$ indeed. The two deterministic attractors $\pm \bar{M}$ with $\bar{M}= \sqrt{\e}\beta \bar{A} = 0.1935$ are highlighted by  horizontal black dotted lines. The small crosses highlight a ``transition''  (see definition in text) of the WNL signal  from the neighborhood of one attractor to the neighborhood of the other.  \label{fig:3}}
\end{figure}

Let us now activate the stochastic forcing ($\phi \neq 0$). We choose the forcing structure to be $\bff = \bq^{\da}$, which maximizes the absolute value of $\eta$ in (\ref{eq:coeffs}), thus the impact of the forcing. We numerically find $\eta = 1.492$. For $\Ren=100$, we show in figure~\ref{fig:3} some temporal signals of $M$ for four gradually increasing forcing amplitudes $\phi$, of the same given noise realization $\xi(\al \e t)$. The cut-off frequency $\om_{co}$ of the slowly-varying noise is set at $\om_{co}=0.04$. Due to the non-normality of the operator $L$, this specific choice is not arbitrary and will be detailed in the last section of the paper. The WNL predictions and the DNS data are compared directly, as they are forced by the exact same noise. 
%%%%%%%%%%%%%%%%%%%
%%%%%%%%%%%%%%%%%%%
%%%%%%%%%%%%%%%%%%%
%%%%%%%%%%%%%%%%%%%
For both signals, under the stochastic forcing, $M$ experiences some random oscillations in the neighborhood of one of the two attractors. After some time, these oscillations may by chance become sufficiently strong such that $M$ transits to the neighborhood of the other attractor, as it overcame the potential barrier separating the two. This scenario is increasingly likely with the forcing amplitude. With the exception of some relatively short episodes, the agreement between WNL and DNS signals is visually excellent in figure~\ref{fig:3}, at least for the considered noise realization. A small but systematic overestimation of the WNL prediction is to be noticed though, already present at the deterministic level and observable in figure~\ref{fig:2b} for $Re=100$. Some discontinuities can also be noticed in the DNS signal, due to the multiplication with the $\sgn$ function in the definition of $M$. 

\begin{figure} 
\centering
  \begin{subfigure}[b]{0.44\linewidth}
\includegraphics[trim={0cm 0.0cm 0cm 0.0cm},clip,width=1\linewidth]{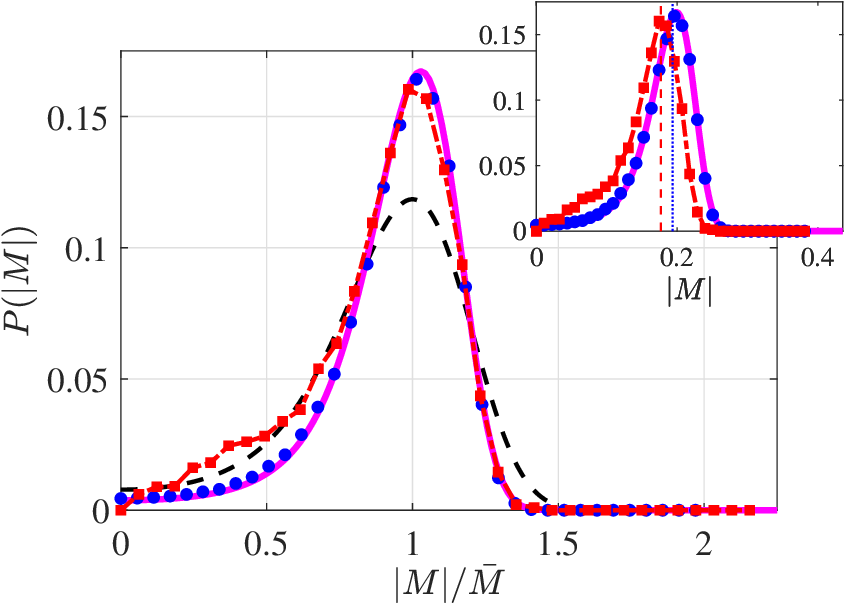}% Here is how to import EPS art
\caption{$\phi = 10$\label{fig:4a}}
\end{subfigure}
%%%%%%%%%%%%%%%%%%%
  \hspace{1cm}
%%%%%%%%%%%%%%%%%%%  
  \begin{subfigure}[b]{0.44\linewidth}
\includegraphics[trim={0cm 0.0cm 0cm 0.0cm},clip,width=1\linewidth]{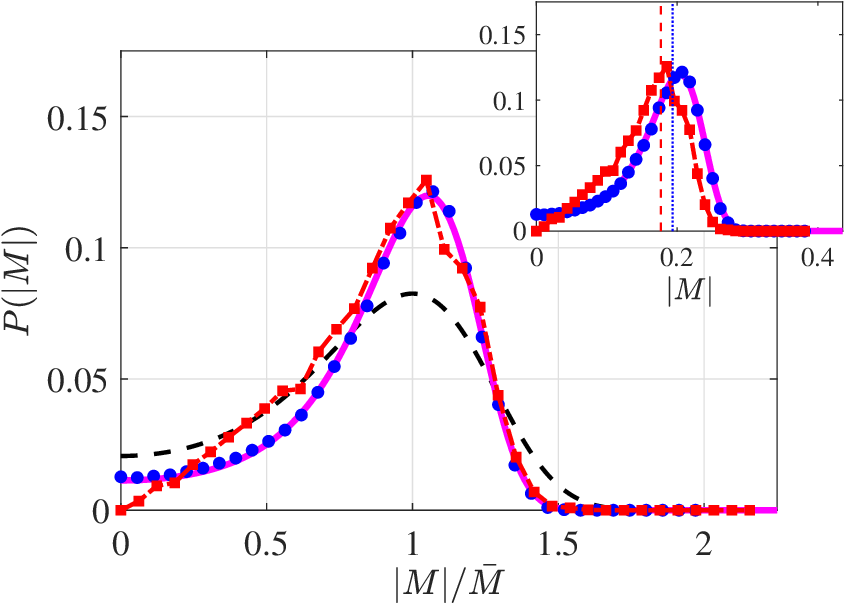}% Here is how to import EPS art
\caption{$\phi = 14$\label{fig:4b}}
 \end{subfigure}
%%%%%%%%%%%%%%%%%%% 
%%
%%%%%%%%%%%%%%%%%%%  
  \begin{subfigure}[b]{0.44\linewidth}
\includegraphics[trim={0cm 0.0cm 0cm 0.0cm},clip,width=1\linewidth]{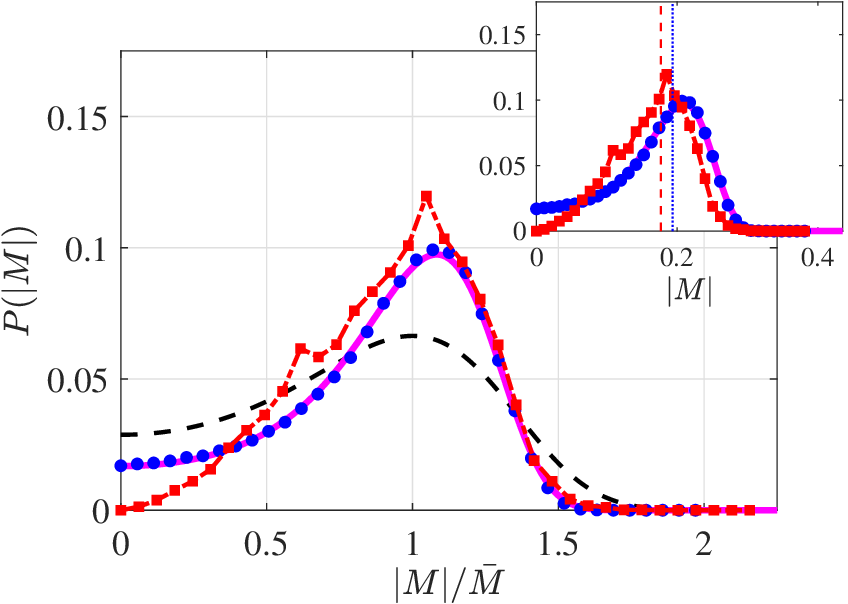}% Here is how to import EPS art
\caption{$\phi = 18$\label{fig:4c}}
 \end{subfigure}
%%%%%%%%%%%%%%%%%%% 
\hspace{1cm}
%%%%%%%%%%%%%%%%%%%  
  \begin{subfigure}[b]{0.44\linewidth}
\includegraphics[trim={0cm 0.0cm 0cm 0.0cm},clip,width=1\linewidth]{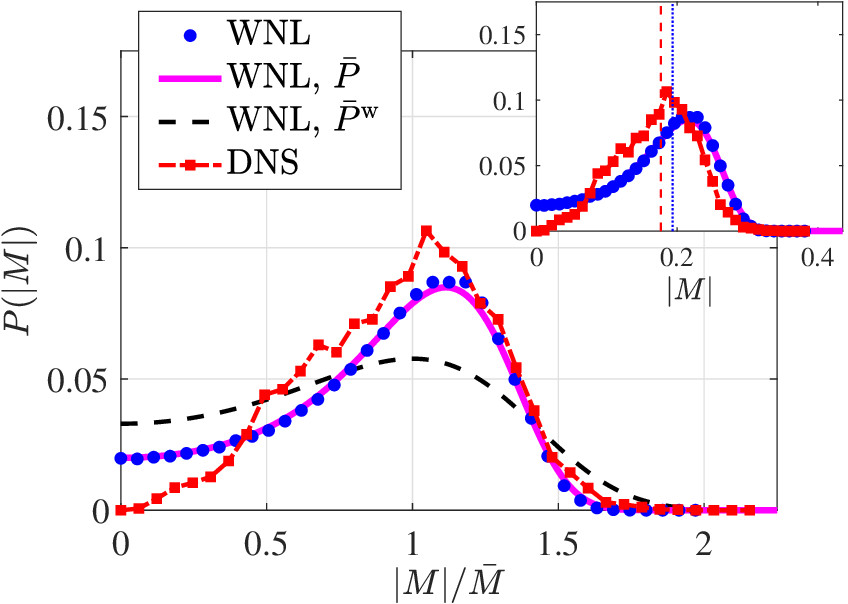}% Here is how to import EPS art
\caption{$\phi = 22$\label{fig:4d}}
 \end{subfigure}
%%%%%%%%%%%%%%%%%%% 
\caption{Probability density function of $|M|$ for the four different forcing amplitudes considered in figure~\ref{fig:3}. The external parameters are also the same as for figure~\ref{fig:3}. For the blue dots ``WNL" and the red dashed-dotted line with the square markers ``DNS", ten simulations, each corresponding to a different noise realization and long of $t=1.1\times 10^4$ units of times, were performed and post-treated. The magenta line ``WNL, $\PS$", defined in (\ref{eq:ps}), is the steady solution of the Fokker-Planck equation (\ref{eq:FP}). The black dashed line ``WNL, $\PSw$", defined in (\ref{eq:PSw}), also corresponds to the steady solution of the Fokker-Planck equation but in a case of ``pure" white noise where $\alpha \om_d \rightarrow \infty$. The insets show to the same data as the main figures, but as a function of $|M|$ instead of $|M|/\bar{M}$. The vertical red dashed line is the deterministic attractor $\bar{M}$ of the DNS whereas the vertical, dotted, blue line is the one from the WNL approach (see figure~\ref{fig:2b}) \label{fig:4} }
\end{figure}

A more quantitative and systematic comparison of both approaches is performed by running, for a given forcing amplitude, nine additional simulations to that of in figure~\ref{fig:3}, each corresponding to a different random noise realization. This generated a sufficient amount of data for convergence of the PDF of $|M|$, shown in figure~\ref{fig:4} for $\phi\in[10,14,18,22]$. We also show $\PS$, the PDF associated with the steady solution of the Fokker-Planck equation (\ref{eq:FP}), defined in (\ref{eq:ps}). Its asymptotic shape in the limit where $\alpha \om_d$ tends to infinity, $\PSw$ defined in (\ref{eq:PSw}), is also visible. 
%%%%%%%%%%%%%%%%%%%
%%%%%%%%%%%%%%%%%%%
%%%%%%%%%%%%%%%%%%%
%%%%%%%%%%%%%%%%%%%
%%%%%%%%%%%%%%%%%%%
%%%%%%%%%%%%%%%%%%%
First, we observe that $\PS$ agrees poorly with $\PSw$, from which we conclude that accounting for a filtered noise in the Fokker-Planck model, at a cut-off frequency $\al \om_d$ (over the slow time, thus $\e \al \om_d$ over the fast), has a significant effect. The probability density function $\PS$ is more localized than $\PSw$ around $\bar{M}$, presumably because the noise corresponding to the former is filtered and thereby has a lower root-mean-square (by Parseval's theorem) than the noise of the latter, thus is less efficient is dislodging $M$ from one of its attractors.

On the other hand, from the excellent agreement between $\PS$ and the PDF obtained from direct simulations of the amplitude equation (\ref{eq:ampeq}), we also conclude that the results are robust to the order of the filter that generates the slow noise; indeed, $\PS$ is associated with a first-order filtered noise whereas the noise in (\ref{eq:ampeq}) is infinite-order filtered (square signal in the frequency domain). Therefore, one does not need to be too careful in the way the slow noise is constructed.

The agreement between $\PS$ and the PDF reconstructed from the DNS is globally satisfactory when plotted over $|M|/\bar{M}$. When plotted over $|M|$ (see inset of figure~\ref{fig:5a}), both PDFs are slightly offset due to the difference between deterministic attractors. In addition, the agreement seems to degrade for $|M| \rightarrow 0$. This is explained by the fact that $M$, due to its definition as a $L^2$ norm of the cross-wise velocity (\ref{eq:M}), is null if and only if the cross-wise velocity is strictly null everywhere along the symmetry axis; because, for instance, of higher-order nonlinear terms neglected in the expansion, this condition is very unlikely to be met in the DNS. This effect is all the more pronounced by increasing $\phi$.

%%%%%%%%%%%%%%%%%%%
%%%%%%%%%%%%%%%%%%%
%%%%%%%%%%%%%%%%%%%
\subsection{Stochastic regime: escape time statistics}
%%%%%%%%%%%%%%%%%%%
%%%%%%%%%%%%%%%%%%%
%%%%%%%%%%%%%%%%%%%

By considering the absolute value $|M|$, the transition events from the neighborhood of one attractor to the other were not considered. However, as developed in the introduction, they are of great interest in practice and thus are studied thereafter. Let us first put a formal definition of what we mean by ``transition". A ``transition" from the neighborhood of one attractor to the other is decreed whenever the following scenario occurs: at some $t_1$, $M$ goes above $-c \bar{M}<0$ (resp. below $c\bar{M}>0$), and doesn't go below (resp. above) this same threshold again before going above (resp. below) the opposite one $c\bar{M}$ (resp. $-c\bar{M}$) at some $t_3$, where the constant $c>0$ is chosen, perhaps arbitrarily, as $c=0.8$; then a transition has occurred at the largest of all time(s) in the interval $[t_1,t_3]$ for which $M$ is null. Under this definition, the transitions are highlighted for the WNL signal by the black crosses in figure~\ref{fig:3}. For $\phi=10$ the time interval $\Delta T$ separating two transitions, shown with an arrow in figure~\ref{fig:3}, sometimes called an ``escape time" or ``first passage time", seems to be on average extremely long. This also underlines the importance of using a reduced-order model. In a simulation long of $\tau=30$ (corresponding to $t=\tau/\e=1.1\times10^4$) units of times, only two transitions could be captured, at a large computational cost for the DNS. By increasing $\phi$ to $\phi=22$, the oscillations of $M$ around $\pm \bar{M}$ are more intense, and the transitions are more frequent (thirteen transitions could be captured). 

\begin{figure} 
\centering
  \begin{subfigure}[b]{0.44\linewidth}
\includegraphics[trim={0cm 0.0cm 0cm 0.0cm},clip,width=1\linewidth]{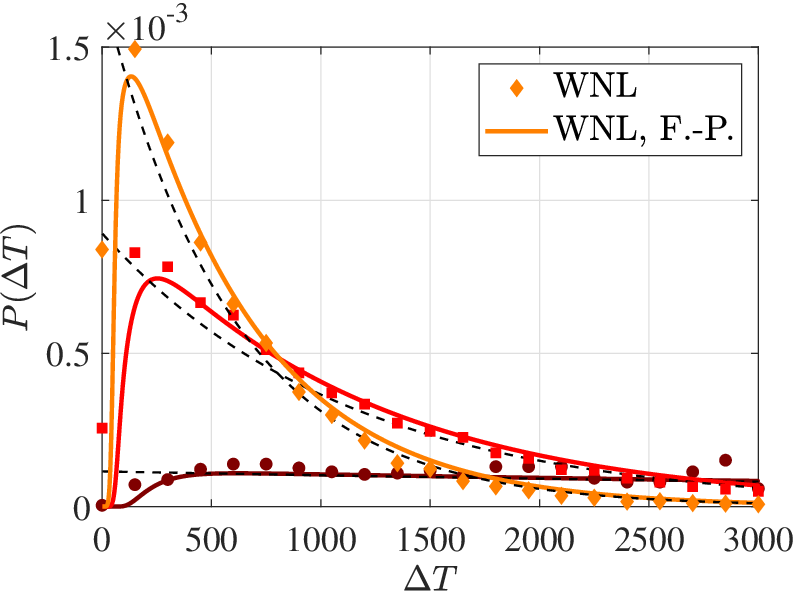}% Here is how to import EPS art
\caption{Lin-lin scale\label{fig:5a}}
\end{subfigure}
%%%%%%%%%%%%%%%%%%%
  \hspace{1cm}
%%%%%%%%%%%%%%%%%%%  
  \begin{subfigure}[b]{0.4475\linewidth}
\includegraphics[trim={0cm 0.0cm 0cm 0.0cm},clip,width=1\linewidth]{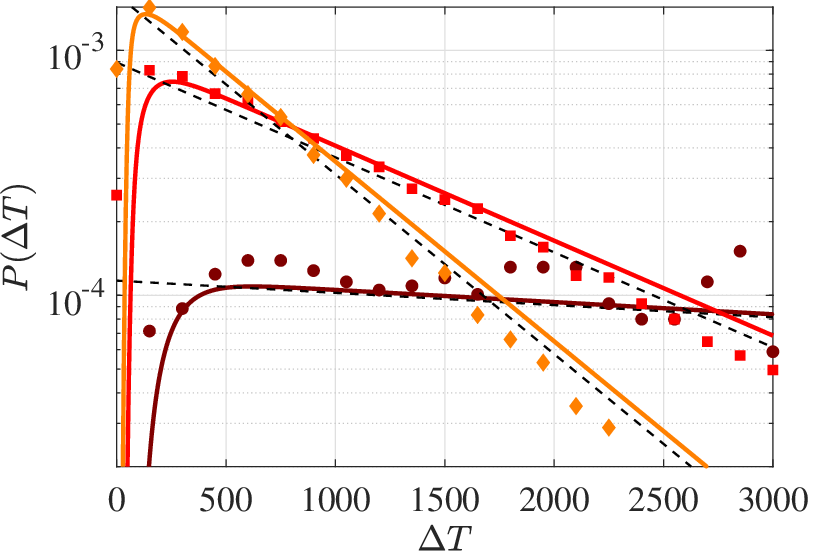}% Here is how to import EPS art
\caption{Log-lin scale\label{fig:5b}}
 \end{subfigure}
%%%%%%%%%%%%%%%%%%% 
\caption{Probability density function of the escape time $\Delta T$ between two transition events from the neighborhood of one attractor to the neighborhood of the other.
Lighter colors correspond to larger forcing amplitude $\phi$ where $\phi \in [10,18,26]$. The parameters are the same as for figure~\ref{fig:3}, specifically $Re=100$ and $\om_{co} = 0.04$. For the markers ``WNL" (dots for $\phi=10$, squares for $\phi=18$ and diamonds for $\phi=26$, ten direct simulations of (\ref{eq:ampeq}), each corresponding to a different noise realization and long of $t=1.1\times 10^6$ units of times, were performed and post-treated. The continuous lines ``WNL, F.-P." (F.-P. for Fokker-Planck),  are obtained by marching in time the Fokker-Planck equation (\ref{eq:FP}) with appropriate initial/boundary conditions (see main text). Subfigures (a) and (b) show the same data, (a) in linear-linear scale and (b) in log-linear scale. The thin black dashed- lines are exponential laws $b\exp{(-b \Delta T)}$, where $b=b(\phi)$ is a fitted parameter. \label{fig:5} }
\end{figure}

In order to approximate the PDF of the escape time $\Delta T$ between two transitions, a sufficiently large number of these transition events has to be reported, and the simulations have to be sufficiently long to also capture very large $\Delta T$, constituting the tail of its PDF. For this reason, for each of the considered forcing amplitude, ten simulations of the amplitude equation (\ref{eq:ampeq}) to an extremely large final time of $\tau=3000$ (corresponding to $t=\tau/\e=1.1\times 10^6$), were performed. Each of these simulations corresponds to a different realization of the slow noise. Some PDF of $\Delta T$ reconstructed by post-processing the so-generated data are proposed in figure~\ref{fig:5} (lin-lin scale in figure~\ref{fig:5a}, and log-lin scale in figure~\ref{fig:5b}). 
%%%%%%%%%%%%%%%%%%%
%%%%%%%%%%%%%%%%%%%
%%%%%%%%%%%%%%%%%%%
%%%%%%%%%%%%%%%%%%%
%%%%%%%%%%%%%%%%%%%
%%%%%%%%%%%%%%%%%%%

Alternatively, the PDF of $\Delta T$ can be computed by marching in time the Fokker-Planck equation (\ref{eq:FP}) with appropriate boundary and initial conditions \cite{Bonciolini18}. The initial condition is set as
%%%%%%%%%%%%%%%%%%%
%%%%%%%%%%%%%%%%%%%
\begin{equation}
\begin{split}
P[A,\xi,\tau=0] = \delta(A-\bar{A}) \frac{\exp\left(-\frac{\xi^2}{2\sigma^2}\right)}{\sigma \sqrt{2\pi}},
\end{split}
\label{eq:FPic}
\end{equation}
%%%%%%%%%%%%%%%%%%%
%%%%%%%%%%%%%%%%%%%
which translates the fact that trajectories are systematically started at the attractor $\bar{A}$, whereas the initial condition for the noise is random and follows a centered normal distribution of standard deviation $\sigma$, i.e. $\xi(t=0)\sim \mathcal{N}(0,\sigma^2)$. Precisely because the noise has zero mean, the variance $\sigma^2$ is equal to the root mean square of the signal, which is expressed by Parseval's theorem
%%%%%%%%%%%%%%%%%%%
%%%%%%%%%%%%%%%%%%%
\begin{equation}
\begin{split}
\sigma^2 \overset{\ea{\xi}=0}{=} \frac{1}{T}\int_{0}^{T}\xi(\al \e t)^2 \di t = \frac{1}{2\pi}\int_{-\infty}^{\infty}|\Fo{\xi(\al\e t)}|^2\di \om = \frac{2\om_{co}}{\e 2\pi} = \frac{\al \om_d}{\pi} = \frac{\al}{\Delta t}
\end{split}
\label{eq:vari}
\end{equation}
%%%%%%%%%%%%%%%%%%%
%%%%%%%%%%%%%%%%%%%
(i.e. two times the area below the red or black curve in figure~\ref{fig:1}, divided by $2\pi$). For the boundary condition, $P=0$ for $|\xi|,A \rightarrow \infty$ is maintained, but instead of also imposing $P=0$ for $A \rightarrow -\infty$, we set $P[-c\bar{A},\xi,\tau]=0$, $\forall \xi, \tau$. As expressed in \cite{Bonciolini18}: ``\textit{this boundary condition is a probability sink, which leads to a monotonic decay in time of the integral} [$\iint P[A,\xi,\tau]\di A \di \xi$, which] \textit{represents the probability of not having crossed the threshold $[-c\bar{A}]$ before time $t$ }. Consequently the PDF of the escape time is the temporal derivative of $1-\iint P[A,\xi,\tau]\di A \di \xi$, for the latter expression is the probability of having escaped to the neighborhood of the attractor $-c\bar{A}$, while (\ref{eq:FPic}) guarantees that all trajectories initially were at the other attractor $\bar{A}$.

The PDFs resulting from this approach are included in figure~\ref{fig:5} and compared with those obtained by post-processing the data generated from direct simulation of (\ref{eq:ampeq}). The agreement between both approaches is globally good. Moreover, it is observed in figure~\ref{fig:5b} that the PDF of the escape time decays exponentially for sufficiently large $\Delta T$; thereby it can be thought of as following an exponential law, reading $be^{-b \Delta T}$. The parameter $b$ is fitted on the PDF obtained with the Fokker-Planck equation, and gives a fair approximation, particularly for the lowest $\phi$ considered in the figure. It is also clear that increasing the forcing amplitude implies a faster exponential decay of the PDF. Indeed, by increasing the intensity of the external excitation, crossing the potential barrier between the two attractors is made easier, thus large escape times are less and less likely.

\begin{figure} 
\centering
  \begin{subfigure}[b]{0.44\linewidth}
\includegraphics[trim={0cm 0.0cm 0cm 0.0cm},clip,width=1\linewidth]{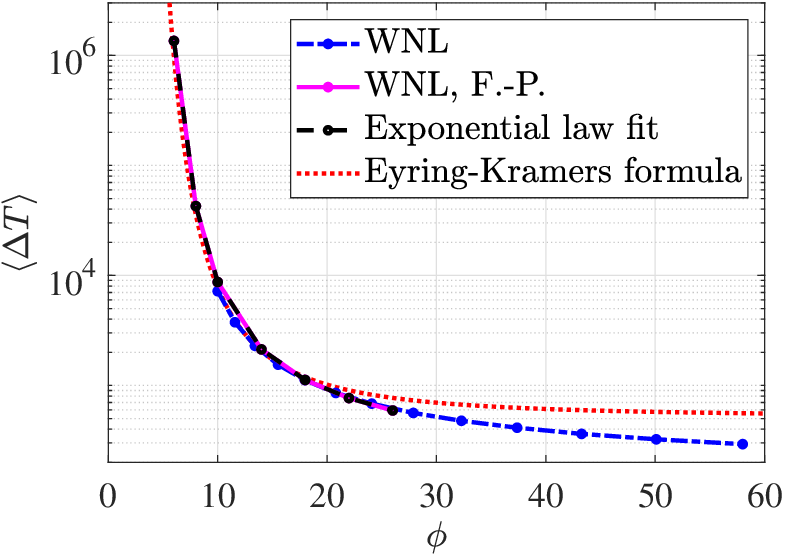}% Here is how to import EPS art
\caption{Mean value of $\Delta T$ \label{fig:6a}}
\end{subfigure}
%%%%%%%%%%%%%%%%%%%
  \hspace{1cm}
%%%%%%%%%%%%%%%%%%%  
  \begin{subfigure}[b]{0.44\linewidth}
\includegraphics[trim={0cm 0.0cm 0cm 0.0cm},clip,width=1\linewidth]{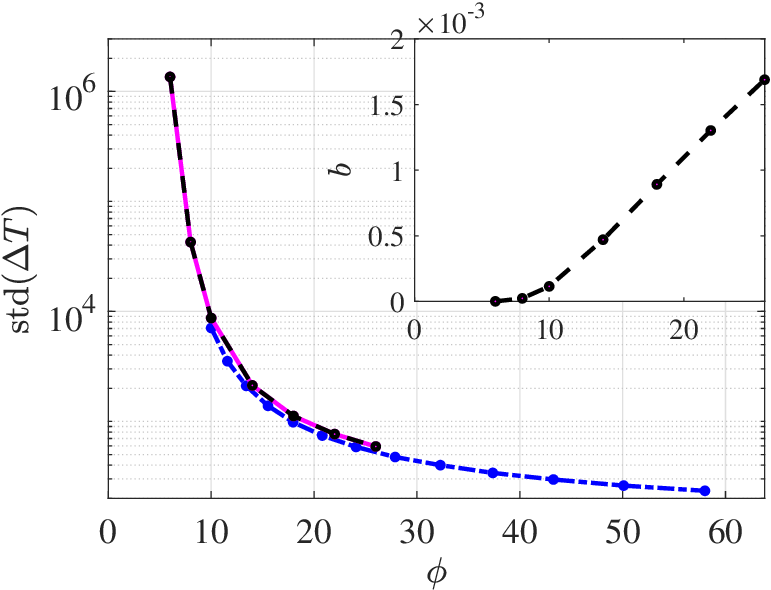}% Here is how to import EPS art
\caption{Standard deviation of $\Delta T$\label{fig:6b}}
 \end{subfigure}
%%%%%%%%%%%%%%%%%%% 
\caption{Mean value and standard deviation of $\Delta T$. The parameters are the same as for figure~\ref{fig:3}, specifically $Re=100$ and $\om_{co} = 0.04$. The blue dash-dotted line is obtained as follows: for each forcing amplitude ten direct simulations of (\ref{eq:ampeq}), each corresponding to a different noise realization and long of $t=1.1\times 10^6$ units of times, were performed and post-treated. The continuous magenta line ``WNL, F.-P." (F.-P. for Fokker-Planck), is obtained by marching in time the Fokker-Planck equation (\ref{eq:FP}) with appropriate initial/boundary conditions. In the subfigure (b) the inset shows the fitted parameter $b$ of the exponential law writing $b\exp{(-b \Delta T)}$. The mean value and the standard deviation reconstructed from this exponential law, both equal to $1/b$, are shown by means of the blacked dashed line. The red dotted line is the Eyring-Kramers formula as given in (\ref{eq:mtt}).\label{fig:6} }
\end{figure}

The mean escape time $\ea{\Delta T}$ and its standard deviation std($\Delta T$) associated with the PDFs in figure~\ref{fig:5} and those for others forcing amplitudes are shown in figure~\ref{fig:6} as a function of the forcing amplitude. 
%%%%%%%%%%%%%%%%%%%
%%%%%%%%%%%%%%%%%%%
%%%%%%%%%%%%%%%%%%%
%%%%%%%%%%%%%%%%%%%
%%%%%%%%%%%%%%%%%%%
%%%%%%%%%%%%%%%%%%%
The agreement between the results from the Fokker-Planck equation and from direct simulations of the amplitude equation is good for the mean escape time. They also collide on the value $1/b$ predicted by the exponential law. Although the differences are barely visible in figure~\ref{fig:6a}, the value $1/b$ is all the closer to $\ea{\Delta T}$ obtained from Fokker-Planck than $\phi$ is small; in other words, the PDF of $\Delta T$ tends towards an exponential law in the limit where $\phi \rightarrow 0$. This result is well-known from the large deviation theory.
%which also predicts the mean escape time to be asymptotic to   

In the ``pure" white noise limit where $\al \om_d \rightarrow \infty$, and for vanishing forcing amplitude, the mean escape time is given by the Eyring-Kramers formula (sometimes also called Kramer's escape rate) according to
%%%%%%%%%%%%%%%%%%%
%%%%%%%%%%%%%%%%%%%
%\begin{equation}
%\begin{split}
%%
%\lim_{\phi \rightarrow 0 } \ea{\Delta T} \propto \frac{1}{\e} \exp \left[-\frac{2\Delta V}{(\eta\phi)^2} \right], \  \text{where} \ \Delta V \doteq V(\bar{A})-V(0) = -\frac{\lambda \bar{A}^2}{4}, 
%%
%\end{split}
%\label{eq:mtt}
%\end{equation}
%%%%%%%%%%%%%%%%%%%
%%%%%%%%%%%%%%%%%%%
%%%%%%%%%%%%%%%%%%%
%%%%%%%%%%%%%%%%%%%
\begin{equation}
\begin{split}
&\lim_{\phi \rightarrow 0 } \lim_{\al \om_d \rightarrow \infty}  \ea{\Delta T} = \frac{1}{\e} \frac{\sqrt{V^{''}(\bar{A})|V^{''}(0)|}}{2\pi} \exp \left[-\frac{2\Delta V}{(\eta\phi)^2} \right],
\end{split}
\label{eq:mtt}
\end{equation}
%%%%%%%%%%%%%%%%%%%
%%%%%%%%%%%%%%%%%%%
where
%%%%%%%%%%%%%%%%%%%
%%%%%%%%%%%%%%%%%%%
\begin{equation}
\begin{split}
\Delta V \doteq V(\bar{A})-V(0) =\frac{\lambda^2}{4\mu}, \quad V^{''}(\bar{A}) = 2\lambda, \quad  \text{and} \quad V^{''}(0)=-\lambda
\end{split}
\end{equation}
%%%%%%%%%%%%%%%%%%%
%%%%%%%%%%%%%%%%%%%
Expression (\ref{eq:mtt}) can be found in \cite{Risken96}, Chapter 5.10, formula (5.111). Without the pre-factor multiplying the exponential, (\ref{eq:mtt}) is referred to as the Arrhenius law in thermodynamics. The relevance of the Eyring-Kramers formula here might appear surprising given the out-of-equilibrium nature of the system (\ref{eq:ns}). Nevertheless, it stems from the fact that, in the specific situation considered in this paper, the Navier-Stokes equation could be reduced to a one-dimensional noisy dynamic, with the deterministic part deriving from a potential. Note that the factor $1/\e$ in (\ref{eq:mtt}) accounts for the fact that the amplitude equation is written over the slow time scale $\tau = \e t$. The Eyring-Kramers formula (\ref{eq:mtt}) is drawn as the red-dotted line in figure~\ref{fig:6a}, and appears accurate until relatively large $\phi\approx 15$. Above this value, it is interesting to notice that the parameter $b$ increases rather linearly with $\phi$, such that the mean escape time decreases as a rational function, thus faster than (\ref{eq:mtt}). 

The evolution of the standard deviation of $\Delta T$ with $\phi$, shown in figure~\ref{fig:6b}, is quantitatively and qualitatively similar to the one of $\ea{\Delta T}$. 

Note that the statistics of the escape time $\Delta T$ shown in figure~\ref{fig:5} and figure~\ref{fig:6} have not been directly compared with those from DNS. That is because, as said, they have been produced over simulations long of $t=1.1\times 10^{6}$ units of time, deliberately extremely long to capture large $\Delta T$. We could not afford DNS of such extreme length, and this is precisely what motivated the approach proposed in this paper. If, as mentioned in the introduction, specific algorithms exist in computing the escape time directly from the fully nonlinear Navier-Stokes equation in (\ref{eq:ns}), their implementation is out of the scope of this paper. Note, however, that a complete comparison between fully and weakly nonlinear escape time statistics would be necessary, since a good agreement between the steady statistics in figure~\ref{fig:4} generically does not imply a good agreement between dynamical quantities such as $\Delta T$. Results shown in figure~\ref{fig:3}, however, suggest that such a comparison would be successful.

%%%%%%%%%%%%%%%%%%%
%%%%%%%%%%%%%%%%%%%
%%%%%%%%%%%%%%%%%%%
\subsection{Stochastic regime: choice of the cut-off frequency}
%%%%%%%%%%%%%%%%%%%
%%%%%%%%%%%%%%%%%%%
%%%%%%%%%%%%%%%%%%%

Let us now say a word about the choice of the cut-off frequency $\om_{co}$. First, it is important to notice that the band-limiting frequency $\om_d = \pi/\Delta t$ can be chosen, in theory, to be arbitrarily large. Therefore, $\om_{co} = \e \al \om_d$, $\e$ being set by the $\Ren$ number and $\alpha=O(1)$, could also be arbitrarily large (as long as it is much smaller than $\om_d$). However, if the linearized operator $L$, defined in (\ref{eq:o1}), is non-normal, there is a specific value of $\om_{co}$ above which we expect the predictions from the amplitude equation to become inaccurate. To determine it, we consider the response of the flow linearized around the neutral, symmetric equilibrium at $\Renc$, $\bU_{\text{c}}$, to a stochastic forcing $\bff \xi(t)$, where $||\bff||=1$ and $\xi(t)$ is a white noise such that $|\hxi(\om)|=1$, $\forall \om$. In the linear paradigm, the amplitude of the forcing term is irrelevant and is set to one for the rest of the reasoning. In the Fourier domain, valid in the limit of large times after the transients fade away, the response writes $\hbu(\om) = R(\om) \bff \hxi(\om)$, the operator $R(\om) \doteq (\ti \om I-L)^{-1}$ being called the resolvent operator. By decomposition on the basis of eigenmodes of $L$, the resolvent operator has a dyadic representation \citep{Luchini14, Schmid14} 
%%%%%%%%%%%%%%%%%%%
%%%%%%%%%%%%%%%%%%%
\begin{equation}
\begin{split}
R(\om) = \sum_{j=1}^{\infty} \frac{1}{\ti\om-\gam_j} \frac{\bq_j \ssp{\bq^{\da}_j}{\ast} }{\ssp{\bq^{\da}_j}{\bq_j}},
\end{split}
\label{eq:Rdyad}
\end{equation}
%%%%%%%%%%%%%%%%%%%
%%%%%%%%%%%%%%%%%%%
where $\bq_j$ (with $||\bq_j||=1$), $\bq^{\da}_j$ (with $||\bq^{\da}_j||=1$) and $\gam_j$ are the $j$th eigenmode, associated adjoint mode and eigenvalue of $L$, respectively. Eigenvalues are ordered such that $\Re(\gam_1) \geq \Re(\gam_2) \geq ...$, implying $\gam_1=0$, $\bq_1 = \bq$ and $\bq_1^{\da} = \bq^{\da}$. The forcing structure considered in the paper was chosen to be $\bff = \bq^{\da}$, which generates 
%%%%%%%%%%%%%%%%%%%
%%%%%%%%%%%%%%%%%%%
\begin{equation}
\begin{split}
R(\om)\bq^{\da} =  \frac{1 }{\ti\om} \frac{\bq}{\ssp{\bq^{\da}}{\bq}} + \sum_{j=2}^{\infty} \frac{1 }{\ti\om-\gam_j} \frac{\bq_j \ssp{\bq^{\da}_j}{\bq^{\da}} }{\ssp{\bq^{\da}_j}{\bq_j}}.
\end{split}
\label{eq:Rdyad2}
\end{equation}
%%%%%%%%%%%%%%%%%%%
%%%%%%%%%%%%%%%%%%%
Had the operator $L$ been normal, both the direct modes and the adjoint modes would form an orthonormal basis and all the inner products $\ssp{\bq^{\da}_j}{\bq^{\da}}$ for $j \geq 2$ would be identically null. This way, the transfer function $||R(\om)\bq^{\da}\hxi(\om)||/||\bq^{\da}\hxi(\om)||=||R(\om)\bq^{\da}||$ reduces to a classical Lorentzian (resonant) response peaked around the resonant frequency $\om=0$
%%%%%%%%%%%%%%%%%%%
%%%%%%%%%%%%%%%%%%%
\begin{equation}
\begin{split}
||R(\om)\bq^{\da}||^2 =  \frac{1}{\om^2\ssp{\bq}{\bq^{\da}}^2}.
\end{split}
\label{eq:Lor}
\end{equation}
%%%%%%%%%%%%%%%%%%%
%%%%%%%%%%%%%%%%%%%
However, $L$ is generally non-normal due to the linearization of the advection term, and neither the direct eigenmodes nor the adjoint ones constitute an orthonormal basis. Therefore the sum in (\ref{eq:Rdyad2}) does not vanish, and the Lorentzian (\ref{eq:Lor}) is only accurate in the limit $|\om|\rightarrow 0$, where the term in $1/(\ti \om)$ in (\ref{eq:Rdyad2}) dominates the sum. 

\begin{figure}
\centering
\includegraphics[trim={0.0cm 0.0cm 0.0cm 0.0cm},clip,width=0.44\linewidth]{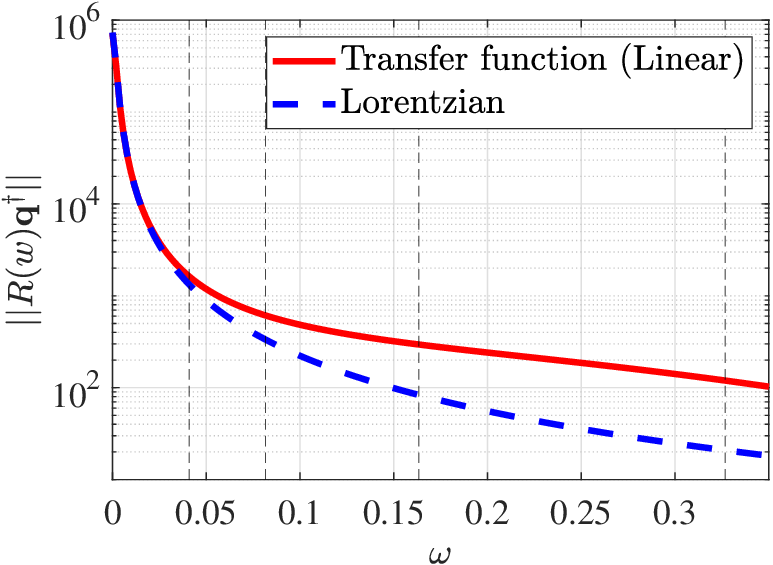}
\caption{The red continuous line is the (linear) transfer function $||R(\om)\bff||$ (norm of the expression (\ref{eq:Rdyad2})) with $\bff = \bq^{\da}$ of the sudden expansion flow linearized around its symmetric neutral equilibrium at $\Ren=\Renc$. The blue dashed line is the Lorentzian function (\ref{eq:Lor}) only accounting for the response of the virtually neutral mode $\bq$, resonant at $\om=0$. Both curves would collide exactly if $L$ was a normal operator, for $\bq^{\da}$ would then excite only $\bq$. Four horizontal black dashed-lines are drawn at $\om \in [4,8,16,32]\times 10^{-2}$.   \label{fig:7}}
\end{figure}

To illustrate this, we compare in figure~\ref{fig:7} the transfer function $||R(\om)\bq^{\da}||$ of the sudden expansion (at $\Ren = \Renc$), with the Lorentzian response (\ref{eq:Lor}).
%%%%%%%%%%%%%%%%%%%
%%%%%%%%%%%%%%%%%%%
%%%%%%%%%%%%%%%%%%%
%%%%%%%%%%%%%%%%%%%
As expected, both curves coincide in the limit $|\om|\rightarrow 0$, where the virtually neutral eigenmode enters in resonance and thus dominates the flow response. Nevertheless, by increasing the frequency above $\om=0.04$, both curves depart from each other and the Lorentzian significantly underestimates the response. This is the consequence of the non-normality of the operator $L$, which implies that $\bq^{\da}$ has a non-null projection over all the other adjoint modes, thus exciting in the response all the associated direct modes; far from the resonant frequency, the eigenmode $\bq$ has no reason to dominate over this response. Consequently, if the noise term contains frequency above $\om_{co}=0.04$, reducing the first-order dynamics of the system on the single mode $\bq$, which was the case in the weakly nonlinear expansion, see (\ref{eq:o1}), might be a poor approximation. 

%However, if the noise term as a cut-off frequency at $\om_{co}=0.04$, the stochastic response remains dominated by $\bq$, such that it is \textit{a priori} justified to reduce the first-order dynamics to the mode $\bq$ only.

\begin{figure} 
\centering
  \begin{subfigure}[b]{0.44\linewidth}
\includegraphics[trim={0cm 0.0cm 0cm 0.0cm},clip,width=1\linewidth]{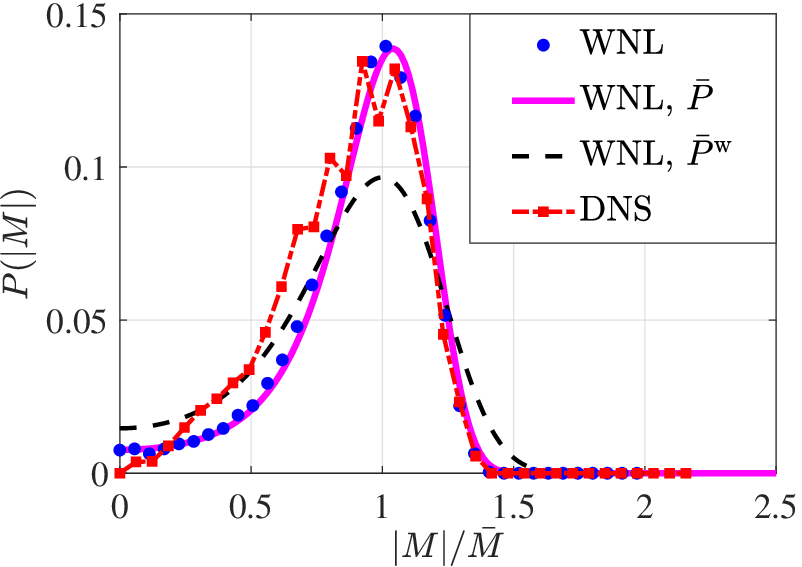}% Here is how to import EPS art
\caption{$\om_{co} = 0.04$ ($\al = 1/8$) \label{fig:8a}}
\end{subfigure}
%%%%%%%%%%%%%%%%%%%
  \hfill
%%%%%%%%%%%%%%%%%%%  
  \begin{subfigure}[b]{0.44\linewidth}
\includegraphics[trim={0cm 0.0cm 0cm 0.0cm},clip,width=1\linewidth]{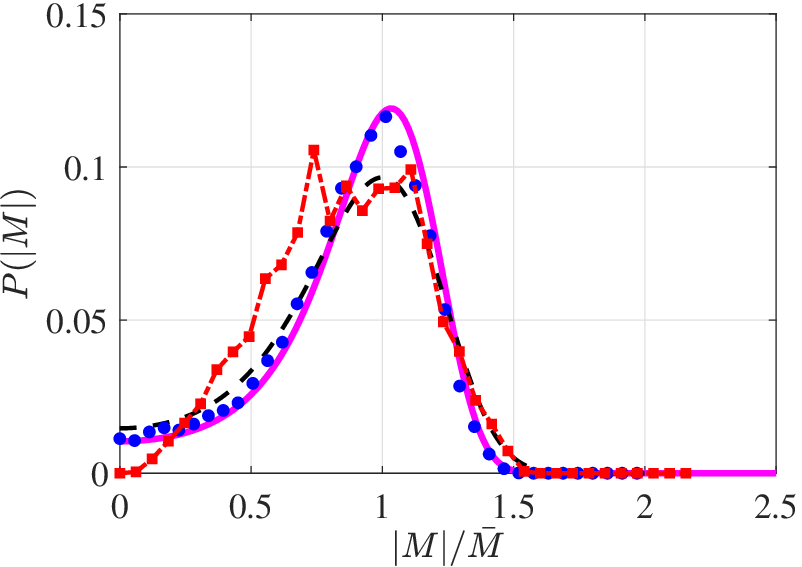}% Here is how to import EPS art
\caption{$\om_{co} = 0.08$ ($\al = 1/4$)\label{fig:8b}}
 \end{subfigure}
%%%%%%%%%%%%%%%%%%%
  \hfill
%%%%%%%%%%%%%%%%%%%  
  \begin{subfigure}[b]{0.44\linewidth}
\includegraphics[trim={0cm 0.0cm 0cm 0.0cm},clip,width=1\linewidth]{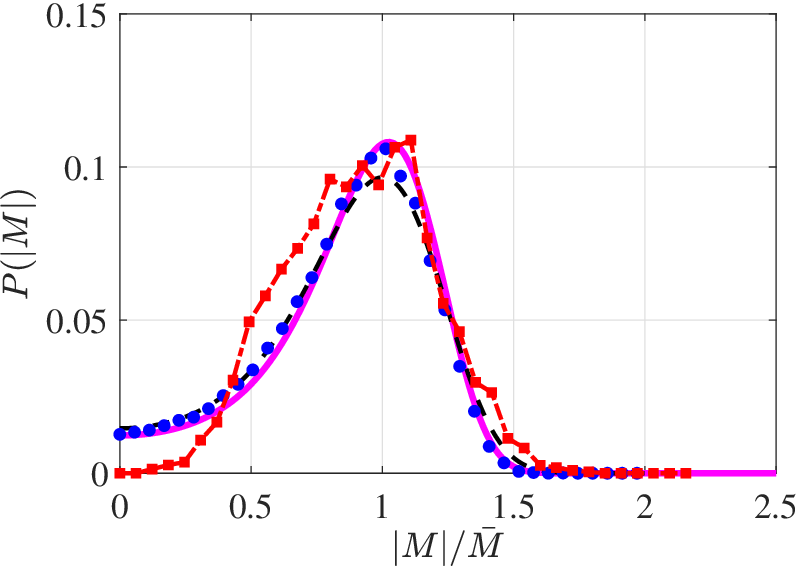}% Here is how to import EPS art
\caption{$\om_{co} = 0.16$ ($\al = 1/2$) \label{fig:8c}}
 \end{subfigure}
%%%%%%%%%%%%%%%%%%% 
  \hfill
%%%%%%%%%%%%%%%%%%%  
  \begin{subfigure}[b]{0.44\linewidth}
\includegraphics[trim={0cm 0.0cm 0cm 0.0cm},clip,width=1\linewidth]{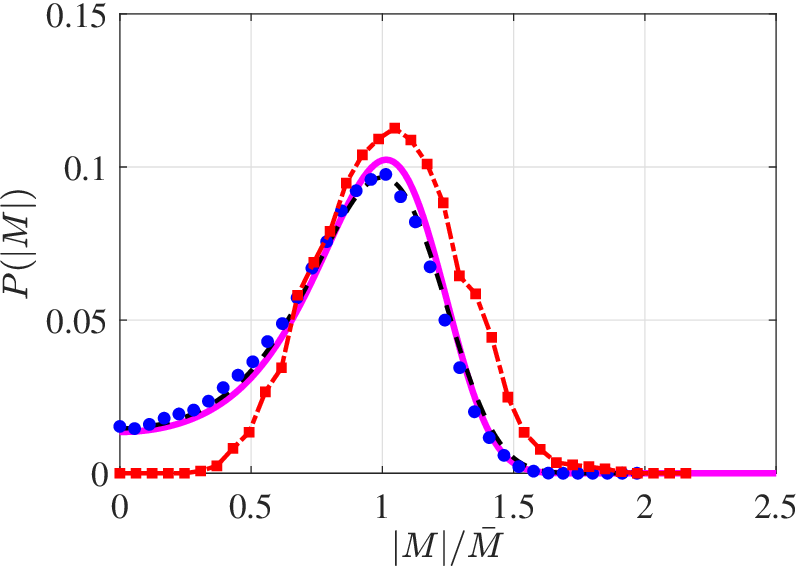}% Here is how to import EPS art
\caption{$\om_{co} = 0.32$ ($\al = 1$)\label{fig:8d}}
 \end{subfigure}
%%%%%%%%%%%%%%%%%%% 
\caption{Probability density function of $|M|$, similarly as in figure~\ref{fig:4} but for a fixed $\phi=12$; to save some computational time, the ten simulations, each corresponding to a different noise realization have been shortened to $t=1.1\times 10^3$ units of times with respect to the computations shown in figure~\ref{fig:4}. The data are shown for four different values of the cut-off frequency $\om_{co} \in [4,8,16,32] \times 10^{-2}$. These four specific frequencies are highlighted with vertical black dashed lines in figure~\ref{fig:7}. \label{fig:8} }
\end{figure}

This is exemplified in figure~\ref{fig:8}, where we show the probability density function of $|M|$, similarly as in figure~\ref{fig:4}, but for a fixed forcing amplitude $\phi=12$ and four increasing values of the cut-off frequency, $\om_{co}\in [0.04,0.08,0.16,0.32]$. The agreement between the PDF obtained from the WNL approach (direct simulation or Fokker-Planck), and the one extracted from DNS, seems to progressively degrade by increasing $\om_{co}$. This is particularly true for $M$ close to zero where the weakly nonlinear PDF largely overestimates the nonlinear one. As explained previously, that is because for too large $\om_{co}$ the eigenmode $\bq$, which is the only one described by the amplitude equation, does not dominate the flow response anymore, and other modes reveal themselves. Due to the activity of these auxiliary modes, the probability of having a null or very low cross-wise velocity along the symmetry axis is reduced, and it becomes more and more unlikely for $|M|$ to take null or small values. Presumably for the same reasons, it appears in figure~\ref{fig:8d} that for the largest considered $\om_{co} = 0.32$, on the contrary, large values of $|M|$ become more likely in the DNS than in the WNL method. As a side comment, note also that increasing $\om_{co}$ incidentally makes $\PS$ converge towards $\PSw$, as expected. 
%%%%%%%%%%%%%%%%%%%
%%%%%%%%%%%%%%%%%%%
%%%%%%%%%%%%%%%%%%%
%%%%%%%%%%%%%%%%%%%
%%%%%%%%%%%%%%%%%%%
%%%%%%%%%%%%%%%%%%%
\section{Summary and perspectives}

In summary, based on the existing literature we have proposed a method to derive a stochastically forced equation for the amplitude of the slowly varying dominant eigenmode, after a steady symmetry-breaking bifurcation. To this purpose, the noise too was  assumed to be slowly varying, in other words, filtered at a certain cut-off frequency  much slower than the band-limiting frequency of the overall system. We gave a precise manner to evaluate \textit{a priori} the cut-off frequency for which the weakly nonlinear expansion, that reduces the linear dynamics to a single eigenmode, is justified. It is the frequency above which the transfer function of the linearized flow departs from a Lorentzian, that only encompasses the dominant mode. We also showed that the order of the filter, generating the slow noise from a classical white noise, mattered little. 

The probability density function of the mode amplitude obtained from the amplitude equation, either by direct simulations or by solving the related Fokker-Planck equation, compared well with the stochastically forced direct numerical simulations. Unsurprisingly, this is particularly true for small forcing amplitude, as increasing the latter increases the relative importance of higher-order nonlinear terms that have been neglected in the weakly nonlinear expansion, and renders small values of the asymmetry measure $M$ unlikely.

The amplitude equation has reduced the dynamics of the flow to a single coordinate whose dynamics derive from a potential. This is particularly convenient when it comes to computing escape time statistics. Indeed, for vanishing forcing amplitude, the waiting time between two events where the solution transits from the neighborhood of one attractor to the other, was found to increase as the exponential of the inverse of the forcing intensity squared. Clearly, this precludes the utilization of direct numerical simulations. On the other hand, as a reduced-order model, the amplitude equation can make predictions at a low numerical cost. 

Nevertheless, even in different cases where the obtained amplitude equation will contain more than one degree of freedom, therefore where generically no potential function exists, the method proposed in this paper can be seen as bridging a system of very high dimension, governed by the Navier-Stokes equation, with the large deviation theory and/or the Fokker-Planck equation that are very effective in systems of low dimension. In this manner, predictions about rare transition statistics could be made without relying on intensive numerical techniques.    

For future research, the method outlined in this paper shall be extended to more general stochastic forcings than a single spatial structure multiplied by a slow noise. Instead, one could consider forcing the flow as in \cite{Farrell93} with a sum of orthonormal forcing spatial structures,  multiplied by uncorrelated white noise processes. As in \cite{Farrell93}, this orthonormal family of forcing structures can be sorted in descending order from largest to smallest maintained variance of the linear response.
Because of the non-normality of the linearized operator, it is possible that only a few of the leading forcings contribute to most of the total variance and thus need to be included. The leading (also called ``optimal") forcing structure will be the adjoint mode, as considered in this paper, for the associated modal response is resonant in $\om=0$ thus its variance (proportional to the integral of (\ref{eq:Lor}) over the frequencies) diverges. The sub-optimal forcing structures, however, will be orthogonal to the adjoint and trigger streamwise convective non-normal amplification in the flow at non-zero frequencies.  At a nonlinear level, the modal and non-normal responses will interact with each other, which is not taken into account in the present analysis, and knowing the role played by the non-normal response at non-zero frequencies over the noise-induced transitions would be of great interest.
For this purpose, the method proposed in \cite{Ducimetiere22a, Ducimetiere22b} to derive amplitude equations for non-normal responses, could be coupled with that proposed in the present paper for the modal one.
In considering noises that are not slowly varying, one could think of splitting slowly and rapidly varying parts and consider the linear response to the rapidly varying part in the Fourier domain at third-order. 

Eventually, we believe that, although concerned with the rather specific configurations of laminar flows past a supercritical bifurcation and subject to external stochastic forcing, the present study could be seen as part of a more general and fundamental study on out-of-equilibrium systems with infinitely many degrees of freedoms. This includes fluid flows in a turbulent regime, where rare transitions are observed in numerous situations, as presented in the introduction. It could be interesting to extend the method proposed here to transitions between turbulent large-scale coherent structures, where the stochastic driving is endogenous and results from nonlinear interactions of the fluctuations. For instance, the construction of a stochastically forced amplitude equation could be done on a turbulent mean flow obtained \textit{a priori} with a quasi-linear analysis as developed in \cite{Farrell12}. Indeed, such turbulent mean flows are also subject to bifurcations and multi-stability, as clearly shown in \cite{Farrell12, Constantinou14, Parker13} and many others. 

The scope also covers other physical systems governed by other stochastic ordinary or partial differential equations. This includes active matter, population dynamics, adaptive networks, microbiological systems, climate science, and many others. As an example, it could be interesting to apply the present method in the spatiotemporal system of bacteria considered in \cite{Grafke17} and subject to a subcritical pitchfork bifurcation above a critical mean bacterial density. Consequently, a stable solution made of a dense bacterial colony appearing at one boundary of the domain co-exists with another symmetrically placed at the other boundary of the domain. It was shown in \cite{Grafke17} that, in certain regimes, the introduction of noise triggers rare and aperiodic transitions between these two solutions. 

%\nocite{*}

\begin{acknowledgments}
The authors wish to thank two anonymous referees, whose comments have substantially improved the manuscript. This work was supported by the Swiss National Science Foundation (grant number 200341).
\end{acknowledgments}

\bibliography{apssamp}% Produces the bibliography via BibTeX.

\end{document}